\documentclass[12pt,a4paper,twoside]{amsart}

\usepackage[T1]{fontenc}
\input xypic\usepackage{amsmath}
\usepackage{amssymb}

\usepackage{amssymb}
\usepackage{latexsym}
\usepackage{graphicx}
\usepackage{psfrag}
\usepackage{times}
\usepackage{mathptm} 

\newtheorem{definition}{Definition}[section]

\newtheorem{lemma}[definition]{Lemma}

\newtheorem{proposition}[definition]{Proposition}

\newcommand{\be}{\begin{equation}}
\newcommand{\ee}{\end{equation}}
\newcommand{\ba}{\begin{eqnarray}}
\newcommand{\ea}{\end{eqnarray}}
\newcommand{\e}{{\rm e}}

\def \bfo {\begin {eqnarray*} }
\def \efo {\end {eqnarray*} }
\def \beq {\begin {eqnarray}}
\def \eeq {\end {eqnarray}}

\def\pmb#1{\setbox0=\hbox{#1}
\kern-.025em\copy0\kern-\wd0
\kern-.05em\copy0\kern-\wd0
\kern-.025em\raise.0433em\box0}

\def \be {\begin{equation}}
\def \ee {\end{equation}}
\def \ba {\begin{eqnarray}}
\def \ea {\end{eqnarray}}

\def \e {\epsilon}

\def \ba {{\bf b}_1}

\def \e {\epsilon}
\def \la {\lambda}

\def \si {\sigma(y)}
\def \sipl {\sigma_+}
\def \simi {\sigma_-}
\def \sidir {\sigma'(y)}
\def \sidpl {\sigma'_+}
\def \sidmi {\sigma'_-}

\begin{document}
\title{Dynamic inverse problem in a weakly laterally inhomogeneous  medium.}
\author{A.S. Blagovestchenskii}\address{A.S. Blagovestchenskii,
Department of Mathematical Physics, St.Petersburg State University, 
St-Petersburg, Russia}

\author{Y. Kurylev}
\address{Yaroslav Kurylev, Department of Mathematical Sciences, Loughborough
Univ.,
Loughborough, LE11 3TU, UK}
\email{Y.V.Kurylev@lboro.ac.uk}

\author{V. Zalipaev}\address{ V.Zalipaev,  Department of Mathematical Sciences, Loughborough
Univ.,
Loughborough, LE11 3TU, UK}
\email{V.Zalipaev@lboro.ac.uk}



\maketitle{Abstract} {\it An inverse problem of wave propagation into a weakly laterally  
inhomogeneous medium  occupying a half-space is considered in the acoustic approximation.  
The half-space 
consists of an upper layer and a semi-infinite bottom separated with an interface. An assumption 
of a weak lateral inhomogeneity means that   the velocity of wave propagation and the 
shape of the interface depend weakly  on the horizontal coordinates, $x=(x_1,x_2)$, in 
comparison  with the strong dependence on the vertical coordinate, $z$, giving rise to a 
small parameter $\e <<1$.   Expanding the velocity in power series with respect to $\e$, 
we obtain a recurrent system of 
 1D inverse problems. We provide algorithms to solve these problems for the zero 
 and first-order approximations.
 In the zero-order approximation,  the corresponding 1D inverse problem is reduced to a 
 system of non-linear Volterra-type integral equations.   In the first-order approximation, 
 the corresponding 1D inverse problem is reduced
to a system of  coupled linear Volterra  integral equations.  These equations are used for the 
numerical reconstruction of the velocity in both layers and the interface up to $O(\e^2)$.
}


\section{Introduction}

The inverse problems of geo-exploration implies investigation of 
domains  inside the earth's crust which contain gas, oil or 
other  minerals  as well as determination of their physical 
properties  such as density, porosity, pressure, and so on.  The most important problem is 
determination of the
boundaries of  the domains which gives  information about its 
location, volume and makes possible  to predict costs 
and outcomes  in exploitation. 
In practice,  in e.g. oil exploration 
and seismology this inverse problem  is mainly solved by means of the 
map migration method. 
The map migration method assumes the knowledge of the velocity profile
above the interface and provides robust, effective numerical algorithms which
are quite stable with respect to  variations of the velocity and complicated geometry
of the interface. Therefore, it has attracted much attention of mathematicians and geophysicists, see 
e.g.   \cite{BCS},
\cite{Kl}, \cite{Mi}, \cite{Wh}, with new approaches continuing to appear e.g. \cite{Kat-Po}.
However, an absence of an independent way to recover the velocity profile above the interface
may hinder the map migration techniques. This makes crucial for geosciences to  develop
algorithms to solve an inverse problem of the velocity reconstruction from the measurements of the
wave field on the ground surface, $z=0$.

 This gives rize to a fully 
non-linear
dynamic multidimensional inverse problem. 
To our knowledge, the only method valid for 
arbitrary  inhomogeneous
velocity profiles is  
 the Boundary Control 
method (BC-method), see e.g. 
 \cite{Be2}, \cite{KKL}. However, at the moment,  this method is developed 
only for smooth  velocity profiles, in the absence of discontinuities. 
Moreover, existing variants of the BC-method  do not possess 
good stability properties and have a rather poor numerical implementation, 
e.g. \cite{Be-Go}. 

 On the other 
hand,   in the case of a pure layered structure, with the properties depending only 
on depth, $z$,  inverse problems of geophysics are often reduced to one-dimensional
inverse problems. The theory of one-dimensional inverse problems which goes back to the 
classical results obtained by Borg, Levinson, Gel'fand-Levitan, Marchenko and Krein of 
late 40-th - 50-th
is still an area of active research with new theoretical results and numerical algorithms
appearing regularly in mathematical and geophysical literature, see e.g. \cite{Bor}, 
  \cite{BuBa}, \cite{Chapman},  \cite{Klyatskin}, 
  \cite{LRY}, \cite{Po}-- \cite{Sy}
to 
mention just a few.

Returning to geophysical applications, a pure layered structure occurs rare. However,
in many cases 
in seismology and oil exploration the parameters of the earth are 
not described 
by arbitrary functions of 3 dimensions having jumps 
across arbitrary 2 dimensional surfaces.
Rather the properties of the 
medium depend mainly on the depth, $z$, with only  
slow dependence 
on
horizontal coordinates, $x=(x_1,x_2)$, i.e. depend on $\e x, \, \e<<1$ rather than $x$ (in seismology,
often $\e \approx 0.1-0.2$) and we deal with a weakly laterally inhomogeneous medium (WLIM).
The importance of WLIM is now well-understood in theoretical and mathematical geophysics. There are currently numerous results on the direct problem of the wave propagation in  WLIM, see e.g.
 \cite{Bche} ,  \cite{Bbul},  \cite{Bbor}, \cite{BST}, \cite{Bly}, \cite{EH}. 
However,
to our knowledge,  except \cite{Bu2}, there is currently no special inversion techniques for WLIM. 
In this paper we develop and numerically implement an inversion method 
for WLIM based on different ideas 
than those in \cite{Bu2}.
This method provides, for inverse problems occurring  in important 
practical applications, a technique which inherits some practically useful properties 
of the one-dimensional inverse problems, e.g.  robustness and fast convergence rate 
of numerical algorithms. 

In the acoustical approximation we use in this paper, the wave propagation in WLIM is described by the wave equation,
\be
u_{tt}-c^2(z,x)\Delta_{z,x} u=0, \quad z>0\label{weq}
\ee
with 
\be
c(z,x)=c_0(z)+\e <x, c_1(z)>+....\label{speed}
\ee
Due to (\ref{speed}), the response of the media, measured at $z=0$, may also be 
decomposed in power series with respect to  $\e$. Analyzing this decomposition, 
we obtain a recurrent system of inverse problems for subsequent terms $c_p(z)$. 
The inverse problem for $c_0(z)$ reduces to a 
family of one-dimensional inverse problems. 
 We employ as a basis to solve this problem  the method coupled non-linear Volterra 
 integral equations developed by Blagovestchenskii, e.g.  \cite{Bl11},  
 which goes back to \cite{Bl2}.  Note that \cite{Bl2}, \cite{Bl11} provide a {\it local} 
 reconstruction of $c_0$ near the ground surface $z=0$. In this paper,
 we pursue the method of coupled non-linear Volterra 
 integral equations further making it {\it global}, i.e. providing  reconstruction of 
 $c_0(y)$ from inverse data measured for $t \in (0,2T)$ up to the depth
 $T$ in the travel-time coordinates, $y$ (see formula (\ref{travel-time})
 and obtain conditional Lipschitz stability estimates for this procedure.
 The results of the zero-order reconstruction  serve as a starting point to develop a 
 recurrent procedure to determine, from the inverse data,  further terms  in 
 expansion (\ref{speed}).
  Such procedure, for the inverse problem of the
 reconstruction of an unknown potential $q(z,\e x)$
rather than $c(z,\e x )$,was suggested by Blagovestchenskii \cite{Bl4}, \cite{Bl11}. 
The method of \cite{Bl11} is based on the use of polynomial moments with 
respect to $x$. 
We suggest another approach based on the
 Fourier transform, $x=(x_1,x_2)\to \xi=(\xi_1,\xi_2)$,
  of the equation (\ref{weq}).
  This makes possible to utilize the dependence of the resulting problems on $\xi$ 
in order to solve the recurrent system of 1D inverse problems. The coefficients $c_1(z), c_2(z),...$ 
may then be obtained as solutions to some linear  Volterra integral equations.
 Having said so, 
we note that the integral equations for higher order unknowns contain higher and higher
 order derivatives of the previously found terms, thus increasing ill-posedness of 
 the inverse problem. This is hardly surprising taking into account  well-known strong 
 ill-posedness of multi-dimensional inverse problems \cite{EnHaNu}, \cite{Ma}. 
 What is, however, interesting is that within the model considered there is just a gradual 
 increase of instability adding two derivations at each stage of the reconstruction algorithm.

In this paper we confine ourself to the   reconstruction only of $c_0,\,c_1$. Reconstruction 
of the higher order terms $c_p,\, p>1$ is, in principle,  possible using the same ideas as 
for $c_1$, although is more technically involved. However, in practical applications in 
geophysics the measured data make possible to find  inverse data only for  $c_0,\, c_1$. 
Indeed, (\ref{weq}), (\ref{speed}) imply that, for the type of the boundary sources we consider,
$$
u(z,x,t)=u_0(z,x,t)+\e u_1(z,x,t)+\e^2 u_2(z,x,t)+...~~,
$$
so that the measured data at $z=0$,
$$
u_z|_{z=0}(x,t)=R(x,t)=R_0(x,t)+\e R_1(x,t)+\e^2 R_2(x,t)+...~~.
$$
What is more, $R_p(x,t)$ are even with respect to $x$ for even $p$ and $R_p(x,t)$ are odd for odd $p$. Clearly, in real measurements $R$ is not given as a power series with respect to $\e$. However, using the fact that $R_0$ is even, and $R_1$ is odd, we can  find $R_0$ and $R_1$ up to $O(\e^2)$ from the measured $R$.


The plan of the paper is as follows. In the next section we give a rigorous formulation of the 
problem  and provide a general outline of the perturbation scheme for this problem in WLIM. 
In  section 3 a modified method of coupled non-linear Volterra integral  equations to 
reconstruct  $c_0(z)$ is described. In  section 4 we derive a coupled 
system of linear Volterra integral equations for $c_1(z)$. 
Section 5 is devoted to the global solvability of the non-linear system for $c_0(z)$ and 
conditional stability estimates. In  section 6 we test the method numerically 
for the two dimensional case (inhomogeneous half-plane).  
As we stop with $c_0,\, c_1$,  in practical applications this would result in an error of the magnitude 
 $O(\e^2)$. The final section is devoted to some concluding remarks.

\section{Formulation of the problem}

Let us consider the wave propagation into the inhomogeneous acoustic half-space 
\be
n^2(z,\e x)u_{tt}-\frac{1}{\rho}{\rm div}(\rho\nabla u)=0,  \quad x=(x_1,x_2),
\ee
where $n$ is the refractive index, $n=c^{-1}$. Above and below the interface $z=h(\e x)$ 
the refractive index is described by
$$
n(z,\e x)=\left  \{ \begin{array} {rcl}   n^{(1)}(z,\e x),&0<z<h(\e x),    \\  n^{(2)}(z,\e x),&z>h(\e x), \end{array}  \right. ,
$$
where $\e$ is a small parameter ($0<\e\ll 1$) which characterizes the ratio of the horizontal and vertical gradients of $n$. We assume that  the density is piece-wise constant,
$$
\rho=\left  \{ \begin{array} {rcl}   \rho_1,&0<z<h(\e x),    \\  \rho_2,&z>h(\e x), \end{array}  \right.
$$
with known $\rho_1,\,\rho_2$. As we deal with WLIM, the following notations are  used,
\be
\label{25.1}
 n^2(z,\e x)=n^2_0(z)+\e <x,\bar n (z)>+O(\e^2), \quad \bar n(z)=(n_1(z),n_2(z)),
\ee
\be
\label{25.2}
h(\e x)=h_0+\e <x, \bar h>+O(\e^2),\quad \bar h=(h_1,h_2),
\ee
where $<,>$ means the scalar product.


We assume that $u=0$ for $t\le 0$, and the boundary condition is 
\be
u{\Bigg |}_{z=0}=\delta(x) f(t)\sqrt{\frac{\rho_1}{n_0(0)}},
\quad f(t)=\delta(t)~~{\rm or} \,\, f(t)=\theta(t),
\ee
where $\delta(x)$ and $\theta(t)$ are $\delta$-function and Heaviside function, correspondingly.
On the interface between these two layers we assume the usual continuity conditions,
\be
\label{interface}
[u]{\Bigg |}_{z=h(\e x)}=0,\quad [\rho^{-1}\frac{\partial u}{\partial n}]{\Bigg |}_{z=h(\e x)}=0,
\ee
where $[\dots]$ stand for a jump across the interface. Using the Fourier transform with respect to $x$
\be
u(z,x,t)=\frac{1}{4\pi^2}\int\limits_{R^2}e^{-i<\xi, x>}U(z,\xi,t)d\xi, \quad \xi=(\xi_1,\xi_2),.
\ee
$U$ can be expanded into  asymptotic series,
\be
U(z,\xi,t) \approx \sum_{n=0}^{\infty}\e^n i^n U^{(n)}(z,\xi,t),
\label{vser}
\ee
where all functions $U^{(n)}(z,\xi,t)$ are real. Using decompositions
(\ref{25.1}), (\ref{vser}), it is easily seen that $U^{(n)}(z,\xi,t)$ are even functions with respect to $\xi$ for even $n$ and odd for  odd $n$.

Our goal is to reconstruct the refractive index and the shape of interface, namely, the functions $n_0(z),\, n_{1,2}(z)$ and constants $h_0, \, h_{1,2}$, from
the response data collected during time $0<t<2T$,
$$
\frac{\partial u}{\partial z}\Bigg |_{z=0}=R(x,t,\e),
$$
with 
$$
R(x,t,\e) \approx \sum_{n=0}^{\infty}\e^n R_n(x,t).
$$
Decomposing the wave equation (\ref{weq}) and interface conditions (\ref{interface}) with respect to $\e$,
 we obtain initial-boundary value problems for  $U^{(0)}$ and $U^{(1)}$. 
The zero-order problem is
\be
\label{0-eq}
n^2_0(z)U^{(0)}_{tt}-U^{(0)}_{zz}+|\xi|^2U^{(0)}=0, \quad |\xi|^2=\xi_1^2+\xi_2^2,
\quad
U^{(0)}\Bigg |_{z=0}=f(t)\sqrt{\frac{\rho_1}{n_0(0)}}
\ee
with the interface continuity conditions 
\be
\label{0-interface}
[U^{(0)}]\bigg |_{z=h_0}=0,\quad
[\frac{1}{\rho}U^{(0)}_z]\bigg |_{z=h_0}=0.
\ee
and inverse data of the form
\be
\label{26.15}
U^{(0)}_z\Bigg|_{z=0}=r_0(t,\xi)=\int\limits_{R^2}\cos{(<\xi, x>)}R_0(x,t)dx.
\ee

The first-order problem is
\be
\label{1-eq}
n^2_0 (z)U^{(1)}_{tt}-U^{(1)}_{zz}+|\xi|^2U^{(1)}=<\bar n,\nabla_{\xi}U^{(0)}_{tt}>,
\quad
U^{(1)}\Bigg |_{z=0}=0.
\ee
with the interface continuity conditions 
\be
\label{1-interface}
\bigg [U^{(1)}-<\bar h,\nabla_{\xi} U^{(0)}_z> \bigg ] \bigg |_{z=h_0}=0,
\quad
\bigg [\frac{1}{\rho}\bigg \{ U^{(1)}_z-<\bar h,\nabla_{\xi} U^{(0)}_{zz}>+U^{(0)}<\xi, \bar h> \bigg \} \bigg ]\bigg |_{z=h_0}=0
\ee
and inverse data of the form
\be
\label{26.16}
U^{(1)}_z\Bigg|_{z=0}=r_1(t,\xi)=\int\limits_{R^2}\sin{(<\xi, x>)}R_1(x,t)dx.
\ee

In this paper we confine ourselves to the  reconstruction of only $n_0,~~\bar n$ and $h_0,~\bar h$. The reconstruction of the higher order terms is, in principal, possible using the same ideas as for $\bar n$ and $\bar h$, although is more technically involved. Moreover, in practical applications in geophysics the measured data make possible to find the inverse data only for  $n_0,~\bar n$ and $h_0,~\bar h$. Using the experimental data of the response $R(x,t)$, one cannot expand it into  power series with respect to $\e$. However, as $R_0$ is even  and $R_1$ is odd with respect to $x$, we have
$$
r_0(t,\xi)=\int\limits_{R^2}\cos{(<\xi, x>)}R(x,t)dx+O(\e^2), \quad
r_1(t,\xi)=\e^{-1}\int\limits_{R^2}\sin{(<\xi, x>)}R(x,t)dx+O(\e^2).
$$
So, the only quantities we may observe from the experiment are $r_0(\xi,t)$ $r_1(\xi,t)$ up to error $\e^2$. Thus, although
the inverse problems for the higher-order terms in (\ref{25.1}) can be considered analytically, inverse data for them  are not available from the measurements.
In this connection,  we do not discuss higher approximations in this paper.
Also the exact value for the parameter $\e$ is not known from the experiment. 
However, this quantity may be 
evaluated numerically using inverse data of response $R(x,t,\e)$. For example, one of 
the ways to calculate $\e$ is as follows
$$
\e=sup\bigg|\int\limits_{R^2}\sin{(<\xi, x>)}R(x,t) dx \bigg|, \quad 0 \le t \le 2T,~~|\xi| \le \xi_{max}, 
$$
where $\xi_{max}$ is a positive  real number of order 1. It is clear that $\e$ is qualitatively related to an extent of the medium being a  weakly laterally inhomogeneous one.

\section{Inverse problem in the zero-order approximation}

\subsection{Half-space}

\noindent In this section we describe an algorithm to solve the inverse problem in the zero-order approximation for an inhomogeneous half-space. Namely, we will describe an algorithm to determine $n_0(z)$ from the response  $r_0(t, \xi)$.
This is a generalization of the approach developed first by Blagovestchenskii \cite{Bl2}, see also \cite{Bl11}. The crucial point of the approach is the derivation of a non-linear Volterra-type system of integral equations  to solve the inverse problem.

Let us introduce two new independent variables 
\be
\label{travel-time}
y=\int\limits_0^z n_0(z)dz,\quad \si=\frac{n_0(z(y))}{\rho}.
\ee
The function $\si$ is called acoustic admittance while $y$ is the vertical travel-time.
Then, 
\be
U^{(0)}_{tt}-\frac{1}{\si}\frac{\partial}{\partial y}\bigg ( \si U^{(0)}_y\bigg)+\frac{|\xi|^2}{n_0^2}U^{(0)}=0, \label{zeron}
\ee
Let $f(t)=\delta(t)$. Then, the  boundary condition and response are
$$
U^{(0)}\bigg |_{y=0}=\delta(t)\sqrt{\frac{\rho_1}{n_0(0)}},~~~~~U^{(0)}_y\bigg |_{y=0}=\frac{r_0(\xi,t)}{n_0(0)}.
$$
Let us change the dependent variable
\be
\psi_1(y,t)=\sqrt{\si}U^{(0)}, \quad \psi_2(y,t)=\frac{\partial \psi_1}{\partial t}+\frac{\partial \psi_1}{\partial y},\label{psi}
\ee
thus reducing the second order PDE for $U^{(0)}$ to a system of two PDE of the first order
\be
\left  \{ \begin{array} {rcl} \psi_{1t}+\psi_{1y}&=&\psi_2      \\    \psi_{2t}-\psi_{2y}&=&q(y)\psi_1, \end{array}  \right.\label{syst1}
\ee
where
\be
q(y,\xi)=-\frac{(\sqrt{\si})''}{\sqrt{\si}}-\frac{|\xi|^2}{n_0^2}.   \label{q}
\ee
Expressions in the left-hand side of (\ref{syst1}) are the total derivatives of $\psi_1$ and $\psi_2$ along corresponding characteristics (see Fig.1). Integrating the first equation along the lower characteristics, and  the second equation - along the upper one, we obtain for $t>y$
\be
\left  \{ \begin{array} {rcl} \psi_1(y,t)&=& \int\limits_0^y \psi_2(\eta,t+\eta-y)d\eta,   \\    \psi_2(y,t)&=&-\int\limits_0^y q(\eta,\xi)\psi_1(\eta,t-\eta+y)d\eta+g(t+y,\xi),  \end{array}  \right.\label{syst2}
\ee
where
\be
g(t,\xi)=\delta'(t)+\frac{n_0'(0)}{2n_0(0)}\delta(t)+\frac{r_0(t,\xi)}{\sqrt{\rho_1n_0(0)}}\label{g}
\ee
is bounded as $t\to 0$ due to the cancellation of  $\delta'(t)-$ and $\delta(t)-$ 
singularities in the right-hand side of (\ref{g}).

\begin{figure}[h]
\begin{center}
\includegraphics[scale=0.4,angle=0.0]{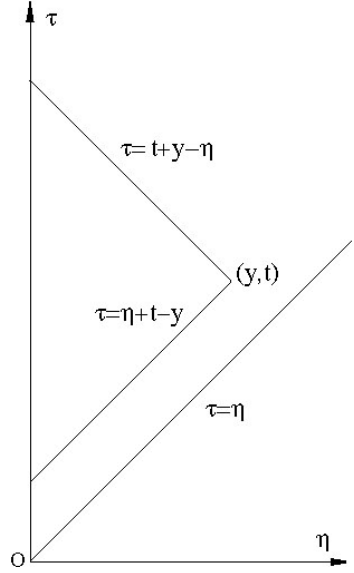}
\end{center}
\label{fig2}
\caption{Integration along characteristic lines: integral for $\psi_1$
is along $\tau=\eta+t-y$, for $\psi_2, \, q$ - along $\tau=t+y-\eta$}
\end{figure}

The system (\ref{syst2}) is  not complete to solve the inverse problem as it has three unknown functions $\psi_1(y,t)$, $\psi_2(y,t)$ and $q(y,\xi)$ and just two equations. We use the progressive wave expansion, see e.g. \cite{Bl11}, \cite{Rom},
\be
\label{25.5}
U^{(0)}(y,t)=\sum_{n=0}f_n(t-y)U^{(0)}_n(y),\quad f_{n+1}(t)=\int\limits_0^t f_n(t)dt, \quad f_0(t)=\delta(t)
\ee 
to derive the third equation.
The amplitudes $U^{(0)}_n$ are found from the transport equations with e.g.
\be
\label{25.6}
U^{(0)}_0=\frac{1}{\sqrt{\si}},
\quad U^{(0)}_1=
\frac{1}{2\sqrt{ \si}} \int\limits_0^y q(\eta)d\eta,
\ee
so that
$$
\psi_2(y,t)=\frac{1}{2}\theta(t-y)q(y)+....
$$
Here and later we often skip the dependence of $q$ and other functions on $\xi$
when our considerations are independent of its value.

Substituting the second equation of (\ref{syst2}) into this relation, and taking $t \to y+0$, we obtain the third desired equation. We summarize the above results in the follwing

\begin{proposition}
\label{prop:1}
Let $\psi_{1,2}(y,t)$ is obtained from the solution  $U^{(0)}(y,t)$ of the wave equation 
(\ref{zeron}) by formula (\ref{psi}). Then $\psi_{1,2}(y,t)$ together with $q(y)$ satisfy the
following system of non-linear Volterra-type equations,
\be
\left  \{ \begin{array} {rcl} \psi_1(y,t)&=& \int\limits_0^y \psi_2(\eta,t+\eta-y)d\eta,   \\    \psi_2(y,t)&=&-\int\limits_0^y q(\eta)\psi_1(\eta,t-\eta+y)d\eta+g(t+y), \\
q(y)&=&-2\int\limits_0^y q(\eta)\psi_1(\eta,2y-\eta)d\eta+2g(2y). \end{array}  \right. \label{nonlin}
\ee
\end{proposition}
This  non-linear Volterra-type system of integral equations, in the sense of Goh'berg-Krein
\cite{G-Kr},
allows us to determine, at least locally,  the function $q(y)$ from  the response $g(t)$ for $0<t<2T$. 
In section 5 we return to the question of the global solvability of system (\ref{nonlin}).  Having solved this system for two different values $\xi_1$ and $\xi_2$, we determine the refractive index in the zero-approximation,
\be
\label{26.1}
n_0(y)=\sqrt{       \frac{|\xi|_2^2-|\xi|_1^2}       {q(y,\xi_1)-q(y,\xi_2)}     },
\ee
which must be used  together with
$$
z=\int\limits_0^y \frac{dy}{n_0(y)}.
$$

The refractive index $n_0(y)$ may also be determined by solving  system (\ref{nonlin}) just for a single value of $\xi$. 
It can be done by integration the differential equation for $w(y)=\sqrt{\si}$ which follows from (\ref{q})
$$
w''+w q(y,\xi)=-\frac{|\xi|^2}{\rho_1^2w^3}.
$$
The general solution of this equation may be represented in the form, see \cite{Bab}.
$$
w(y)=\sqrt{\sum_{i,j=1,2}A_{i,j}w_i(y)w_j(y)},
$$
where $w_{1,2}(y)$ make a fundamental system of the corresponding homogeneous equation.
Constant real symmetric matrix $A_{i,j}$ satisfies
$$
{\rm det}||A_{i,j}||W^2(w_1,w_2)=-\frac{|\xi|^2}{\rho^2_1},
$$
where
$$
W(w_1,w_2)=w_1w_2'-w_1'w_2.
$$
Given $w(0)$ and $w'(0)$  we are then able to determine  all entries of the matrix $A$.

\subsection{Two layers problem}

\noindent Consider a wave propagation through the interface, $y=y(h_0)=L$,  using the  singularity analysis of the incident, reflected and transmitted waves. Then, in the layer $ 0<y<L$,
the singularities of $U^{(0)}$ given by the incident and reflected waves, are
\bfo
& &U^{(0)}=\frac{\delta(t-y)}{\sqrt{\si}}+d_0\frac{\delta(t+y-2L)}{\sqrt{\si}}+
\\
\nonumber
& &+\frac{\theta(t-y)}{2\sqrt{\si}}Q_0(y)+\frac{\theta(t+y-2L)}{2\sqrt{\si}}(d_1+Q_1(y))+..., \quad
0<y<L, \,\, L<t<2L.
\efo
For the transmitted wave, the WKB asymptotics is given by
$$
U^{(0)}=s_0\frac{\delta(t-y)}{\sqrt{\si}}+\frac{\theta(t-y)}{2\sqrt{\si}}(s_1+Q_1(y))+..., \quad y>L,
\,\,L<t<2L,
$$
where
$$
Q_0(y)=\int\limits_0^yq(\eta)d\eta, \quad Q_1(y)=\int\limits_L^yq(\eta)d\eta.
$$
Unknown quatities $d_0$, $d_1$ and $s_0$, $s_1$ are the reflection and transmission coefficients.
Using the interface continuity conditions (\ref{0-interface}),
leads to a linear system of  equations for $d_0$, $d_1$ and $s_0$, $s_1$ so that 
\beq
\label{reflection}
& &
d_0=\frac{1-\frac{\sipl}{\simi}}{1+\frac{\sipl}{\simi}}, \quad 
d_1=\frac{  \sidmi(1+d_0)+Q_0(L)(\simi -\sipl)-s_0\sidpl\sqrt{\frac{\simi}{\sipl}}      }{\sipl+\simi},
\\
\nonumber
& &
s_0=\sqrt{\frac{\sipl}{\simi}}\frac{2}{1+\frac{\sipl}{\simi}}, \quad
s_1=\sqrt{\frac{\sipl}{\simi}}\frac{\sidmi(1+d_0)+2Q_0(L)\simi   -s_0\sidpl\sqrt{\frac{\simi}{\sipl}}}{\sipl+\simi},
\eeq
where $\sigma_{\pm}$ and $\sigma'_{\pm}$ are the limiting values of $\si$
and $\sidir$ at $y\to L-0$ and $y\to L+0$, correspondingly.

\begin{figure}[h]
\begin{center}
\includegraphics[scale=0.4,angle=0.0]{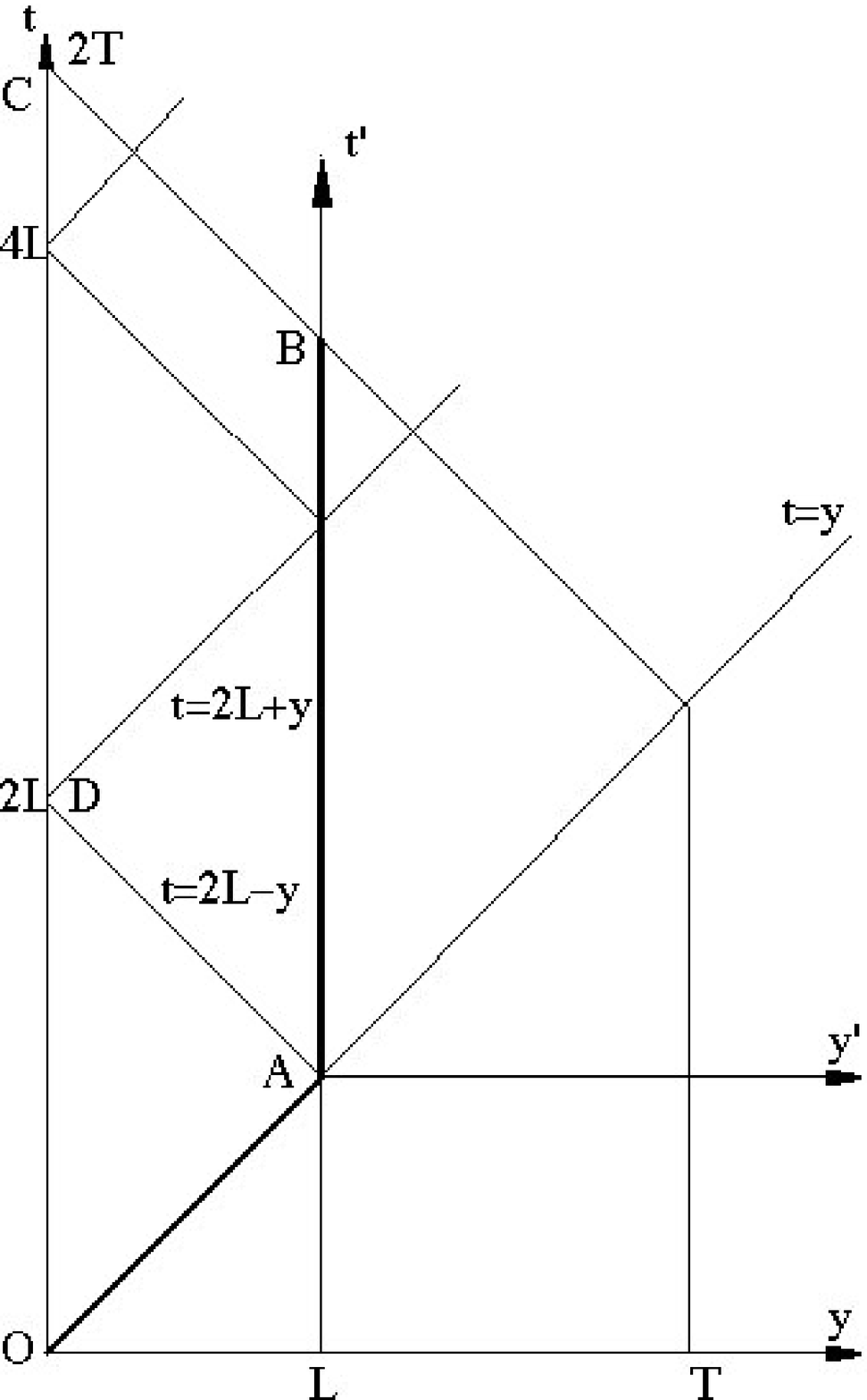}
\end{center}
\label{outinv}
\caption{Characteristic lines in the case of two layers.}
\end{figure}

Analyzing the incoming singularities at $y=0$, we can find $L$ and, therefore,
$h-0$ and also $d_0$ and  $d_1$. This will give us $\sigma_+$ and $\sigma'_+$.
The next step is  reconstruction of the velocity at  large depth  $L<y<T$, which is illustrated on Fig. 2. This reconstruction requires the data observed for  $0<t<2T$. At this stage we again apply the non-linear Volterra system of integral equations (\ref{nonlin}) in the new frame $y', \eta', t'$ ($y=y'+L,\,\eta=\eta'+L,\,t=t'+L$)
\be
\label{new}
\left  \{ \begin{array} {rcl} \psi_1(y',t')&=& \int\limits_0^{y'} \psi_2(\eta',t'+\eta'-y')d\eta'+\psi_1^+(0,t'-y'),   \\    \psi_2(y',t')&=&-\int\limits_0^{y'} q(\eta')\psi_1(\eta',t'-\eta'+y')d\eta'+\psi_2^+(0,t'+y'), \\
q(y')&=&-2\int\limits_0^{y'} q(\eta')\psi_1(\eta',2y'-\eta')d\eta'+2\psi_2^+(0,2y').
\end{array}  \right.
\ee
These require  knowledge of $\psi_1^+(0,t'-y')$ and $\psi_2^+(0,t'+y')$ at the vertical segment $AB$ as the  limits $\psi_{1,2}^+(0,t)=\lim \psi_{1,2}(y,t)$ for $y \to L+0$.  To this end, we
first determine $\psi^-_{1,2}(0,t)= \lim \psi_{1,2}(y,t)$ for $y \to L-0$  using  system (\ref{syst2}).
Employing the interface continuity conditions
$$
\frac{\psi^-_1}{\sqrt{\simi}}=\frac{\psi^+_1}{\sqrt{\sipl}}, \quad
\psi^-_{1y}\sqrt{\simi}-\psi^-_1\frac{\sidmi}{2\sqrt{\simi}}=\psi^+_{1y}\sqrt{\sipl}-\psi^+_1\frac{\sidpl}{2\sqrt{\sipl}},
$$
we obtain the required data, $\psi_1^+(0,t'-y')$ and $\psi_2^+(0,t'+y')$ at  $AB$.

\section{ Inverse problem in the first-order approximation} 

\subsection{Half-space}
In this section we describe an algorithm to  determine 
$
\bar n(z)$
in an inhomogeneous half-space from the response  $r_1(t,\xi)$. For convenience, we  integrate all the functions of the zero-order problem with respect to time $t$ so that 
$$
U^{(0)}\bigg |_{y=0}=\theta(t)\sqrt{   \frac{\rho_1}{n_0(0)}     }.
$$
Using the variables  $y$ and $\si$, 
we rewrite the  problem  (\ref{1-eq}), (\ref{1-interface}) in the form
\be
U^{(1)}_{tt}-\frac{1}{\si}\frac{\partial}{\partial y}\bigg ( \si U^{(1)}_y\bigg)+\frac{|\xi|^2}{n_0^2}U^{(1)}=\frac{1}{n_0^2}<\bar n,\nabla_{\xi}U^{(0)}_{tt}>,\label{pr1}
\ee
$$
U^{(1)}\bigg |_{y=0}=0, \quad U^{(1)}_y\bigg |_{y=0}=\frac{\hat r_1(t,\xi)}{n_0(0)},
\quad \hat r_1(t,\xi)=\int\limits_0^t r_1(t,\xi)dt.
$$
We start with the progressive wave expansion for $U^{(1)}$. As by (\ref{25.5}), (\ref{25.6}),
$$
<\bar n,\nabla_{\xi}U^{(0)}_{tt}>=-<\bar n,\xi>\frac{\delta(t-y)}{\sqrt{\si}}\int\limits_0^y \frac{d\eta}{n_0^2(\eta)}+...~~.
$$
the progressive wave expansion for $U^{(1)}(y,t)$ has the form
$$
U^{(1)}(y,t)=\theta(t-y)A_0(y)+(t-y)_+A_1(y)+....
$$
Therefore, 
 equation (\ref{pr1}) implies, in particular,  that 
\be
\label{26.6}
A_0(y)=\frac{1}{\sqrt{\si}}\int\limits_0^y <\bar n(\eta),\xi>p(\eta)d\eta, \quad p(y)=-\frac{1}{2n_0^2(y)}\int\limits_0^y \frac{d\eta}{n_0^2(\eta)}.
\ee
Let   $G(y,\eta,t;\xi)$ be the space-causal Green's function,
\be
G_{tt}-\frac{1}{\si}\frac{\partial}{\partial y}\bigg ( \si G_y\bigg)+\frac{|\xi|^2}{n_0^2}G=\delta(t)\delta(y-\eta),\quad G=0 \,\, {\rm if} \,\,y<\eta. \label{green}
\ee
Then
\beq
& &U^{(1)}(y,y+0;\xi)=
\\
\nonumber
& & \int\limits_0^yd\eta\int\limits_{\eta}^{2y-\eta}d\tau G(y,\eta,y-\tau;\xi)\frac{<\bar n,\nabla_{\xi}U^{(0)}_{tt}(\eta,\tau;\xi)>}{n_0^2(\eta)}
-\frac{1}{n_0(0)}\int\limits_0^{2y}d\tau G(y,0,y-\tau;\xi)\hat r_1(\tau,\xi),
\eeq
where we now write down explicitly the dependence on $\xi$.

Introducing  new  unknown functions,
\be
\label{26.7}
\bar \varphi(y)=( \varphi_1(y), \varphi_2(y))=\bar n (y) p(y)
\ee
and employing the progressive wave expansion for $U^{(1)}$, we see that
$$
<\bar \varphi(y),\xi>=\frac{d}{dy}\Bigg \{ \sqrt{\si}\int\limits_0^yd\eta\int\limits_{\eta}^{2y-\eta}d\tau G(y,\eta,y-\tau,\xi)\frac{<\bar \varphi(\eta),\nabla_{\xi}U^{(0)}_{tt}(\eta,\tau,\xi)>}{n_0^2(\eta)p(\eta)}
$$
\be
   -\frac{\sqrt{\si}}{n_0(0)} \int\limits_0^{2y}d\tau G(y,0,y-\tau,\xi)\hat r_1(\tau,\xi)\Bigg \}.\label{intrel}
\ee

Finally, the desired system of linear Volterra  integral equations may be obtained from  (\ref{intrel}) 
by differentiation and  setting $\xi$ equal to e.g. $\xi_1=(a,0)$ and  $\xi_2=(0,a)$, where 
$a\neq 0$. Namely, we obtain
\begin{proposition}
\label{prop:2}
Let $\bar \varphi(y)= (\varphi_1(y), \varphi_2(y))$ be given by (\ref{26.7}) where the scalar factor $p(y)$ depends only on the already found $n_0(y)$. Then $\bar \varphi(y)$ satisfies the system of
linear Volterra equations
\bfo
\nonumber
& &
a\varphi_i(y)=2\int\limits_0^yd\eta G_1(y,\eta,\eta-y,\xi_i)\frac{<\bar \varphi(\eta),\nabla_{\xi}U^{(0)}_{tt}(\eta,2y-\eta,\xi_i)>}{n_0^2(\eta)p(\eta)}
\\
\nonumber
& &
 -\frac{2}{n_0(0)}G_1(y,0,-y,\xi_i)\hat r_1(2y,\xi_i)
+\int\limits_0^yd\eta\int\limits_{\eta}^{2y-\eta}d\tau G_2(y,\eta,y-\tau,\xi_i)\frac{<\bar \varphi(\eta),\nabla_{\xi}U^{(0)}_{tt}(\eta,\tau,\xi_i)>}{n_0^2(\eta)p(\eta)}  -
\\
& &-\frac{1}{n_0(0)} \int\limits_0^{2y}d\tau G_2(y,0,y-\tau,\xi_i)\hat r_1(\tau,\xi_i),
\quad i=1,2.
\efo
where 
$$
G_1(y,\eta,t-\tau;\xi)=\sqrt{\si}G(y,\eta,t-\tau;\xi) \quad G_2(y,\eta,t-\tau;\xi)=G_{1t}(y,\eta,t-\tau;\xi)+G_{1y}(y,\eta,t-\tau;\xi).
$$
\end{proposition}
Observe that, due to the first equation in (\ref{nonlin}),
$$
\lim_{\eta\to 0}\nabla_{\xi}U^{(0)}_{tt}(\eta,\tau)=0,
$$
so that the system of integral equtions in Proposition \ref{prop:2}  is not singular.

\subsection{Two layers problem} 

Let us consider the inverse problem for the first-order approximation  in the case of a layer 
and a semi-infinite bottom separated by an interface. Our goal is to derive a coupled system 
of linear Volterra integral equations similar to that in Proposition \ref{prop:2} 
 to 
reconstruct $\bar n(z)$ and also to  determine constants $\bar h$ characterizing this interface. Recall 
the formulation of the corresponding problem using $y$ variable, see (\ref{1-eq}), (\ref{1-interface}),
\be
U^{(1)}_{tt}-\frac{1}{\si}\frac{\partial}{\partial y}\bigg ( \si U^{(1)}_y\bigg)+\frac{\xi^2}{n_0^2}U^{(1)}=\frac{1}{n_0^2}<\bar n,\nabla_{\xi}U^{(0)}_{tt}>, \label{inheq}
\ee
$$
U^{(1)}\bigg |_{y=0}=0, \quad U^{(1)}_y\bigg |_{y=0}=\frac{\hat r_1(t,\xi)}{n_0(0)}.
$$
$$
\bigg [U^{(1)}-n_0<\bar h,\nabla_{\xi} U^{(0)}_y> \bigg ] \bigg |_{y=L}=0,
$$
$$
\bigg [ \si U^{(1)}_y-\bigg (\rho\sigma^2(y)<\bar h,\nabla_{\xi} U^{(0)}_{yy}>+\rho\si \sigma'(y)<\bar h,\nabla_{\xi} U^{(0)}_{y}>-U^{(0)}\frac{<\xi, \bar h>}{\rho}\bigg) \bigg ] \bigg |_{y=L}=0.
$$
As before, we start with the  singularity analysis of the transmitted and reflected waves. The progressive wave expansion of the inhomogeneous term in (\ref{inheq}) is given by
$$
\frac{\partial}{\partial \xi _i}U^{(0)}_{tt}=-\frac{\xi_i\delta(t-y)}{\sqrt{\si}}\int\limits_0^y \frac{d\eta}{n_0^2(\eta)}+\frac{\delta(t+y-2L)}{\sqrt{\si}}\bigg (\frac{1}{2}\frac{\partial}{\partial \xi _i}d_1-\xi_i\int\limits_L^y \frac{d\eta}{n_0^2(\eta)}\bigg )+...~~,
$$
$$
0<y<L,\,\,L<t<2L,\,\,i=1,2,
$$
$$
\frac{\partial}{\partial \xi _i}U^{(0)}_{tt}=\frac{\delta(t-y)}{\sqrt{\si}}\bigg (\frac{1}{2}\frac{\partial}{\partial \xi _i}s_1-\xi_i\int\limits_L^y \frac{d\eta}{n_0^2(\eta)}\bigg )+...~~,\quad y>L,\,\,L<t<3L.
$$
Then the sum of the incident and reflected waves is given by
\be
\label{jump}
U^{(1)}=\frac{\theta(t-y)}{\sqrt{\si}}\int\limits_0^y <\bar n,\xi>p(\eta)d\eta+\frac{\theta(t+y-2L)}{\sqrt{\si}}\bigg (\int\limits_L^y <\bar n,\bar p_1>d\eta+D \bigg)+...~~,
\ee
$$
0<y<L,\,\, L<t<2L,
$$
where
$$
\bar p_1^{(i)}(y)=\frac{1}{2n_0^2(y)}\bigg (\frac{1}{2}\frac{\partial}{\partial \xi _i}d_1-\xi_i \int\limits_L^y \frac{d\eta}{n_0^2(\eta)}\bigg ),~~~~~
0<y<L.
$$
The transmitted wave is then
\be
U^{(1)}=\frac{\theta(t-y)}{\sqrt{\si}}\bigg (\int\limits_{L}^y <\bar n,\bar p_2>d\eta+S\bigg)+..., 
\quad y>L,\,\, L<t<3L, \label{aswkb}
\ee
where
$$
\bar p_2(y)=\frac{1}{2n_0^2(y)}\bigg (\frac{1}{2}\frac{\partial}{\partial \xi _i}s_1-\xi \int\limits_L^y \frac{d\eta}{n_0^2(\eta)}\bigg ),
\quad y>L
$$
and coefficients $d_1, s_1$ are defined in (\ref{reflection}).

Substituting these  expressions for $U^{(1)}$ into the  interface continuity relations
in (\ref{inheq}), and solving the corresponding  system of linear equations, we find that
\beq \label{D}
& &
D(\xi)=\frac{ \bigg (1-\frac{\sipl}{\simi}\bigg )\int\limits_0^L <\bar n,\xi>p(y)dy+\frac{\sipl}{\simi}<\bar h,\bar \Gamma_1>+<\bar h,\bar \Gamma_2>}{1+\frac{\sipl}{\simi}},
\\
\nonumber
& &
S(\xi)=\frac{2\int\limits_0^L <\bar n,\xi>p(y)dy+<\bar h,\bar \Gamma_2-\bar \Gamma_1>}{\sqrt{\frac{\simi}{\sipl}}+\sqrt{\frac{\sipl}{\simi}}}.
\eeq
Here the  vectors $\bar \Gamma_1$ and $\bar \Gamma_2$ are given by
\bfo
& &
\bar \Gamma_1^{(i)}=\rho_1\simi\bigg ( \frac{\partial}{2\partial \xi _i} d_1+\xi_i \int\limits_0^L \frac{d\eta}{n_0^2(\eta)}\bigg )+\sqrt{\frac{\simi}{\sipl}}\rho_2\sipl
\frac12\frac{\partial}{\partial \xi _i} s_1,
\\
\nonumber
& &
\bar \Gamma_2^{(i)}=\rho_1\simi\bigg ( \frac{\partial}{2\partial \xi _i} d_1-\xi_i \int\limits_0^L\frac{d\eta}{n_0^2(\eta)}\bigg )-\sqrt{\frac{\sipl}{\simi}}\rho_2\sipl 
\frac12 \frac{\partial}{\partial \xi _i} s_1, \quad i=1,2.
\efo
From the jump of $U^{(1)}$ 
 at $y=0$ at the time $t=2L$, we can find $D$. 
 Now equation (\ref{D}) with $\xi_1=(a,0)$ and $\xi_2=(0,a)$ make possible to find $\bar h$,
\be
\label{h_1}
h_i=\bigg[D(\xi_i)\bigg (1+\frac{\sipl}{\simi}\bigg )-\bigg (1-\frac{\sipl}{\simi}\bigg )\int\limits_0^L <\bar n,\xi_i> p(y)dy \bigg]
 \bigg [ 2a\rho_1\frac{\sipl-\simi}{\sipl+\simi}  \int\limits_0^L \frac{d\eta}{n_0^2(\eta)}   \bigg ] ^{-1}.
\ee
Clearly, at this stage we can also find $S$.

 Let us describe an algorithm to  determine 
$
\bar n(z)$ in the second layer.
The boundary conditions $U^{(1)}|_{y=L+0}$ and the response $U^{(1)}_{y}|_{y=L+0}$ for 
the transmitted wave can be found using $\hat{r}_1(t, \xi)$ and already known, in the 
upper layer,  $n_0(y),\, \bar{n}(y)$ together with the interface continuity conditions in 
(\ref{inheq}).
 Using the  coordinates $y', \eta', t',\, y=y'+L, \eta=\eta'+L,\,\,t=t'+L$, see Fig. 1, we have the progressive wave 
 expansion for $U^{(1)}$,
$$
U^{(1)}(y',t')=\frac{\theta(t-y')}{\sqrt{\sigma(y')}}\bigg (\int\limits_0^{y'} <\bar n,\xi>p_2(\eta)d\eta+S\bigg)+..., \quad \bar p_2(y')=\xi p_2(y').
$$
Employing the Green function  (\ref{green})  in the second layer, we obtain
\bfo
& &
p_2(y')<\bar n(y'),\xi>=
\\
\nonumber
& &
\frac{d}{dy'}\Bigg \{ \sqrt{\sigma(y')}\bigg (\int\limits_0^{y'}d\eta'\int\limits_{\eta'}^{2y'-\eta'}d\tau G(y',\eta',y'-\tau)\frac{<\bar n(\eta'),\nabla_{\xi}U^{(0)}_{tt}(\eta',\tau,\xi)>}{n_0^2(\eta')}
\\
\nonumber
& &
   -\int\limits_0^{2y'}d\tau G(y',0,y'-\tau)\left(\chi_2(\tau)+\frac{\sidpl}{\sipl}\chi_1(\tau)\right)  + \int\limits_0^{2y'}d\tau G_{\eta'}(y',0,y'-\tau)\chi_1(\tau) 
 \\
 \nonumber
 & &
  + [G(y',0,y')\chi_1(0)-G(y',0,-y')\chi_1(2y')] \bigg ) \Bigg \},
\efo
where 
\bfo
\chi_1(t)= U^{(1)}|_{y=L}, \quad
\chi_2(t)= U^{(1)}_y|_{y=L}
\efo
Finally, the desired coupled system of linear Volterra  integral equations determining $\bar n(y)$ 
below the interface may be obtained from this equation  by setting e.g. $\xi_1=(a,0)$ and  
$\xi_2=(0,a)$  (compare with Proposition \ref{prop:2}). It is worth noting that these integral 
equations are not singular. 

\section{Global reconstruction and stability}
As mentioned in section 3, the non-linear Volterra-type system of integral equations (\ref{nonlin}) 
may,
in principle, be solvable only locally. In this section we show that, assuming {\it a priori} bounds 
for $q(t)$ in $C(0,T)$, it can be found on the whole interval $(0,T)$ using a layer-stripping method 
based on (\ref{nonlin}). What is more, we will prove Lipschitz stability of this procedure with 
respect to the variation of the response data $g(t)$ in $C(0,2T)$.

We first note that if  $||q||_{C(0,T)} < M$ then solving the direct problem gives
\be\label{24.1}
||\psi_{1,2}||_{C(\Delta_T)}, \,\, ||q||_{C(0,T)} < A=A(T,M),
\ee
where we can assume $A \geq \max(1,M)$.
Here $\Delta_T$ and generally
 $\Delta_b,\, 0<b<T$, is  the triangle in $R^2$ bounded by the characteristics 
 $\eta=\tau,\, \eta+\tau=2b$ and the axis $\eta=0$. Assume that we have already found 
 $\psi_{1,2}$ in $\Delta_b$ and $q$ in $(0,b)$. 
 \begin{lemma}
\label{lem:layer-stripping}
Let $\la>0$ satisfies
\be
\label{24.2}
\la < \frac{1}{16(1+A)^2}.
\ee
Then system (\ref{nonlin}) uniquely determines $\psi_{1,2}(y,t), \, q(y)$ for $(y,t) \in \Delta_{b+\la}$,
$y \in (0,b+\la)$, respectively. Moreover, these functions can be found from system (\ref{nonlin})
by Picard iterations.
\end{lemma}

{\bf Proof} Utilizing new coordinates $y'=y-b,\, \eta'=\eta-b,\, t'=t-b$
(compare with the end of section 3), we obtain a system of non-linear Volterra-type
equations similar to (\ref{new}),
\be
\label{syst5}
\left  \{ \begin{array} {rcl} \psi_1(y',t')&=& \int\limits_0^{y'} \psi_2(\eta',t'+\eta'-y')d\eta'+g_1(y',t'),   \\    \psi_2(y',t')&=&-\int\limits_0^{y'} q(\eta',\xi)\psi_1(\eta',t'-\eta'+y')d\eta'+g_2(y',t'), \\
q(y')&=&-2\int\limits_0^{y'} q(\eta')\psi_1(\eta',2y'-\eta')d\eta'+g_3(y').
\end{array}  
\right.
\ee
Here
$
g_1(y',t')=\psi_1(0,t'-y'),\quad
g_2(y',t')=\psi_2(0,t'=y'),
\quad g_3(y')= 2 \psi_2(0,2y')
$
may be found from the "direct" 
system (\ref{syst2}) and satisfy $|g_i|<A$.

Using the first equation in (\ref{syst5}),  we eliminate $\psi_1$ to get
\be
\label{newsystem}
\left  \{ \begin{array} {rcl}    
\psi_2(y',t')&=&-\int\limits_0^{y'} q(\eta',\xi)\left[g_1(\eta',t'-\eta'+y')
+\int_0^{\eta'}\psi_2(\xi',t'-\eta'+\xi') d\xi'\right]d\eta'
+g_2(y',t'), \\
q(y')&=&-2\int\limits_0^{y'} q(\eta') \left[g_1(\eta',2y'-\eta')
+\int_0^{\eta'}\psi_2(\xi',y'-\eta'+\xi') d\xi'\right]
d\eta'+g_3(y').
\end{array}  
\right.
\ee

 We rewrite this system in the  form,
 $$
 (\psi_2,\,q)=K_{\la}(\psi_2,\,q),
 $$
 where $K_{\la}$ is a non-linear operator in $C(\Delta_{\la} \times C(0,\la)$ determined by the right-hand side (\ref{newsystem}).  Let us show that $K_{\la}$ is a contraction in
$$
{\bf D}_{2A}= B_{2A}(C(\Delta_{\la})) \times B_{2A}(C(0, \la)),
$$ 
where $B_{\rho}(\cdot)$ is the ball of radius $\rho$ in the corresponding function space and we use the norm
$$
||(\psi_2,\,q)||= \max(||\psi_2||,\,||q||).
$$
Considering
$$
K_{\la}(\psi_2,q)-K_{\la}({\hat \psi}_2,{\hat q})=\left(K_{\la}(\psi_2,q)-K_{\la}(\psi_2,{\hat q})\right)
+\left(K_{\la}(\psi_2,{\hat q})-K_{\la}({\hat \psi}_2,{\hat q})\right),
$$
it follows from (\ref{newsystem}) that in ${\bf D}_{2A}$
$$
||K_{\la}(\psi_2,q)-K_{\la}({\hat \psi}_2,{\hat q})|| \leq
2A\la \left ( ||q-{\hat q}||+2\la ||\psi_2-{\hat \psi}_2||
+2 \la ||q-{\hat q}|| \right).
$$
As $\la$ satisfies (\ref{24.2}), this implies that
$$
||K_{\la}(\psi_2,q)-K_{\la}({\hat \psi}_2,{\hat q})|| \leq
\frac38 ||(\psi_2,\,q)- ({\hat \psi}_2,{\hat q})||.
$$
Similar arguments show that $K_{\la}:{\bf D}_{2A} \rightarrow {\bf D}_{2A}$. Thus (\ref{newsystem})
has a unique solution in ${\bf D}_{2A}$ and, since (\ref{new}) is a Volterra-type system, in
$C(\Delta_{\la}) \times C(0,\la)$.

\noindent QED

Next we prove the Lipschitz stability of the non-linear Volterra system (\ref{nonlin}). Let us denote
by $\tilde\psi_{1,2}(y,t)$, $\tilde q(y,\xi)$ and $\tilde g(t)$ small variations of the corresponding functions.  
\begin{lemma}
\label{stability}
Let $\psi_{1,2},\, q$ satisfy (\ref{24.1}). There exist $c_0=c_0(T,A),\, \e_0=\e-0(T,A)$ such that for
$
||\tilde g ||_{C(0,2T)} < \e <\e_0
$
the corresponding system (\ref{nonlin})  with $g+\tilde g$ instead of $g$ 
has a unique solution $\psi_{1,2}+\tilde\psi_{1,2}$, $q+\tilde q$,
 and
\be
\label{estimate}
||\tilde\psi_{1,2}||_{C(\Delta_T)},\,\, ||\tilde q||_{C(0,2T)} < c_0 \e.
\ee
\end{lemma}

{\bf Proof}
Substituting expressions for $\psi_{1,2}(y,t)+\tilde\psi_{1,2}(y,t)$, 
$q(y)+\tilde q(y)$ and $g(t)+\tilde g(t)$ into  (\ref{nonlin}) , 
 we obtain the corresponding system in 
variations
$$
\left  \{ \begin{array} {rcl} \tilde\psi_1(y,t)&=& \int\limits_0^y \tilde\psi_2(\eta,t+\eta-y)d\eta,   \\    \tilde\psi_2(y,t)&=&-\int\limits_0^y q(\eta)\tilde\psi_1(\eta,t-\eta+y)d\eta
\\ & & -\int\limits_0^y \tilde q(\eta)(\psi_1(\eta,t-\eta+y)+\tilde\psi_1(\eta,t-\eta+y))d\eta+
\tilde g(t+y), \\
\tilde q(y)&=&-2\int\limits_0^y  \tilde q(\eta)\psi_1(\eta,2y-\eta)d\eta-2\int\limits_0^y  (q(\eta)+
\tilde q(\eta))\tilde\psi_1(\eta,2y-\eta)d\eta+2\tilde g(2y). \end{array}  \right. \label{varnonl}
$$
Let $||\tilde g||<\e$. Then, we have
$$
\left  \{ \begin{array} {rcl} |\tilde\psi_1(y,t)|&\le& \int\limits_0^y |\tilde\psi_2(\eta,t+\eta-y)|d\eta,   \\    |\tilde\psi_2(y,t)|&\le&2A\int\limits_0^y |\tilde\psi_1(\eta,t-\eta+y)|d\eta+A\int\limits_0^y |\tilde q(\eta)|d\eta+\e, \\
|\tilde q(y)|&\le&4A\int\limits_0^y  |\tilde q(\eta)|d\eta+2A\int\limits_0^y  |\tilde\psi_1(\eta,2y-\eta)|d\eta+2\e. \end{array}  \right. 
$$
Substituting the upper inequality into the second one to replace $|\tilde\psi_1(\eta,t-\eta+y)|$, we obtain
$$
|\tilde\psi_2(y,t)|\le 2A\int\limits_{\Delta_{y,t}} |\tilde\psi_2(\eta,\tau)|d\eta d\tau+
2A\int\limits_0^y |\tilde q(\eta)|d\eta+\e,
$$
where $\Delta_{y,t}$ is the the triangle in $R^2$ bounded by the characteristics $\tau=\eta+t-y,\, \tau=t+y-\eta$ and the axis $\eta=0$.
Similarly,
$$
|\tilde q(y)|\le 4A \int\limits_0^y  |\tilde q(\eta)|d\eta+4A\int\limits_{\Delta_{y,t}} |\tilde\psi_2(\eta,\tau)|d\eta d\tau+2\e.
$$
  Let
$$
p(\eta)={\rm max}_{\tau} |\tilde \psi_2(\eta,\tau)|, \quad (\eta, \tau) \in \Delta_T.
$$
Then, we have
$$
|\tilde\psi_2(y,t)|\le 4AT\int\limits_0^y p(\eta)d\eta+2A\int\limits_0^y |\tilde q(\eta)|d\eta+\e,
$$
$$
|\tilde q(y)|\le 4A\int\limits_0^y  |\tilde q(\eta,\xi)|d\eta+8AT\int\limits_0^yp(\eta)d\eta+2\e,
$$
as
$$
\int\limits_{\Delta_{y,t}} p(\eta)d\eta d\tau=2\int\limits_0^yp(\eta)(y-\eta)
d\eta\le 2T\int\limits_0^yp(\eta)d\eta.
$$
Moreover, it holds that
$$
p(y)\le 4AT\int\limits_0^y p(\eta)d\eta+2A \int\limits_0^y |\tilde q(\eta)|d\eta+\e.
$$
Introduce
$$
\rho (y)=p(y)+|\tilde q(y)|.   
$$
Adding the last two inequalities for $p(y)$ and $|\tilde q(y)|$ yields
$$
\rho(y)\le \int\limits_0^y (12ATp(\eta)+6A |\tilde q(\eta)|)d\eta+3\e,
$$
or, with $C=6A{\rm max}(1,2T)$,
$$
\rho (y)\le C\int\limits_0^y
\rho (\eta)d\eta+3\e\le C\int\limits_0^y\bigg(C\int\limits_0^\eta \rho (\eta_1) d\eta_1+3\e \bigg)d\eta +3\e=
$$
$$
\frac{C^2}{1!}\int\limits_0^y \rho (\eta)(y-\eta)d\eta+(Cy+1)3\e.
$$
Continuing this process, we come to the estimate
$$
\rho (y)\le\frac{C^{n+1}}{n!}\int\limits_0^y \rho (\eta)(y-\eta)^nd\eta+\bigg(y^n\frac{C^n}{n!}+y^{n-1}\frac{C^{n-1}}{(n-1)!}+...+1\bigg)3\e.
$$
Let now $
\rho_0={\rm max} \rho (y)$ for $y\in [0,T]$. Then, 
$$
\rho (y)\le \rho_0 \frac{C^{n+1}y^{n+1}}{(n+1)!}+3e^{Cy} \e,
$$
i.e.
$$
\rho_0 \le \rho_0 \frac{(CT)^{n+1}}{(n+1)!}+3e^{CT} \e.
$$
Clearly, for sufficiently large $n$ it holds that
$$
\frac{(CT)^{n+1}}{(n+1)!}\le \frac{1}{2},
$$
and then,
$$
\rho_0 \le 6e^{CT} \e.
$$
Thus, assuming $||\tilde g||\le \e$, we obtain that
\be
\label{10.5.1}
||\tilde q||\le 6e^{CT}\e, \quad ||\tilde \psi_2||\le 6e^{CT}\e, \quad ||\tilde \psi_1||\le 6Te^{CT}\e,
\ee
which implies (\ref{estimate}) with 
$$
c_0=6\max(1,T)e^{CT}, \quad
\e_0= \frac{1}{6\max(1,T)}e^{-CT} \left(A- \max(||\psi_{1,2}||,\, ||q||)\right).
$$

\noindent QED

Observe that when $\tilde g =O(1)$ estimate (\ref{10.5.1}) does not imply, for sufficiently large $T$,
that $\tilde \psi_{1,2}+\psi_{1,2},\, {\tilde q}+q$ 
satisfy (\ref{24.1}). This is a manifestation of a possible 
non-convergence of the Picard method for large $\tilde g$.

Observe also that the conditional stability, in $C(0,T)$,  
of the inversion method based on the Volterra-type system
(\ref{nonlin}) implies the conditional stability, in $C^2(0,T)$, of the original inverse problem
of the reconstruction of $n_0(y)$. Indeed,  if $n_0(y)$ 
is {\it a priori} bounded in $C^2(0,T)$,
i.e. $q(y,\xi)$ is {\it a priori} bounded in $C^2(0,T)$ for bounded $\xi$, 
the inverse map, $g(t,\xi) \rightarrow n_0(y)$, is Lipschitz stable from $C(0,2T)$ to $C(0,T)$.
Due to (\ref{q}), this implies the Lipschitz stability of the above map from $C(0,2T)$ to $C^2(0,T)$.

Regarding  Volterra system in Proposition \ref{prop:2}, we first note that, if $n_0 \in C^3(0,T)$,
then $\triangledown_{\xi} U^{(0)}_{tt} \in C(\Delta_T)$. This implies the Lipschitz stability of this
system  in $C(0,T)$. It is clear from (\ref{26.6}), (\ref{26.7}) 
and can be also checked directly
using the fact that 
$$
{\hat r}_1(t)=
n_0(0) \int \int U^{(0)}(\eta, t-\tau) n_0^{-2}(\eta) < {\bar n}(\eta),\, 
\triangledown_{\xi} U^{(0)}_{tt}(\eta,\tau> d\eta d\tau,
$$
that ${\hat r}_1(t)=t f(t),\, f \in C(0,2T)$. Therefore, the inversion method to find
${\bar n}$ using Proposition \ref{prop:2} is Lipschitz stable into $C(0,T)$ with respect
to the variation of ${\hat r}_1$ in the norm $||t^{-1}{\hat r}_1||_{C(0,2T)}$.

\section{Numerical results} 
In this section we demonstrate  efficiency of the described  method in
numerical solution  of the inverse problem. The computational results  were obtained 
for the  two-dimensional problem with coordinates $(z,x)$. 
In this case,
$$
n^2=n^2_0(z)+\e x n_1(z)+O(\e^2),\quad h(\e x)=h_0+\e xh_1+O(\e^2).
$$
The constant $h_0$ is determined by the arrival time of the $\delta-$type singularity reflected
from the interface, while
 $h_1$ may be evaluated by measuring,  at $z=0$, the amplitudes of 
the reflected   waves, see (\ref{h_1}). In the present numerical implementation,  we assume that 
 their 
values are known.
The described algorithms for solving the zero-order and first-order inverse problems are 
implemented 
into a computer code to reconstruct
 $n_0(z)$ and $n_1(z)$. The first part of the computer code 
generates responses for both orders approximations $r^{(0)}(t,\xi)$ and $r^{(1)}(t,\xi)$. 
The chosen profiles are described by 
$$
\sigma_0(z)=p_0+p_1z+p_2 z^2+q\sin f_0z
$$
for the zero-order problem, and by the trigonometric polynomial 
$$
n_1(z)=r_0+r_1\cos f_1z+r_2 \cos 2f_1z+q_1\sin f_1z+q_2\sin 2f_1z 
$$
for the first-order problem.
Taking various coefficients in the above representations below and above
 the interface, we present on Fig.3 and 4
 the numerically computed values of $\sigma_0(z)$ and $n_1(z)$  against original data.   
 Here we chose $\rho_1=1$ and $\rho_2=1.5$. As we can see there is a good agreement 
 of exact and computed profiles in both cases.

On Fig.5 we demonstrate the results in the case of the response data are corrupted by 
a $2\%$ noise   for $R_0$ and a $10\%$ noise  for $R_1$. The ratio of  2 and 10 may be 
explained by assuming that $\e=\frac{1}{5}$. The data with $5\%$ noise  for $R_0$ and 
$25\%$ for $R_1$ are presented in Fig.6. To clarify the results of numerical reconstructions  one should take into the account that, with respect to the complete refractive index, the error in the first-order approximation should be multiplied by $\e\ll1$.

The method has shown to be quite stable, fast and accurate. When solving Volterra-type integral equations, both non-linear and linear, the iteration processes need just a few iterations (for all graphs the number of iterations was chosen 10).  Clearly, the number of iterations and accuracy in the reconstruction depend on  the scale  of  discretization.  On Fig. 7 we demonstrate the error dependence on the scale of discretization. 

Numerous computer experiments have shown that for a better accuracy and fast convergence 
of the iteration process it is reasonable to use for the chosen profiles the segment $|\xi|<0.5$. 
It is worth noting that the parameter $|\xi|$, the maximum depth $T$ and the maximum of  
$n''_0(z)$ are interconnected. For example, for the larger values of $T$ and the maximum 
of $n''_0(z)$, while computing the profiles we were forced to take smaller values of 
$|\xi|$. 
Moreover, due to the non-linearity of (\ref{nonlin}), when we increase $T$ and/or $n''_0(z)$ 
and $|\xi|$, a blow up effect can occur, i.e. the iterations stop to converge. This may be remedied,
using the results of section 5, 
by a variant of the layer-stripping as it was used in the sections 3.2 and 4.2 but for an interface 
without discontinuity. 

In applications to geophysics, the unity of the refractive index corresponds to the average speed 
of the wave propagation  $c=$2-2.5km/sec. Thus, the dimensionless depth coordinate $z$ 
must be multiplied by $2-2.5 km$.

\begin{figure}[t]
\begin{picture}(400,320)
\put(-20,305){(a)}
\put(-20,120){\includegraphics[scale=0.30]{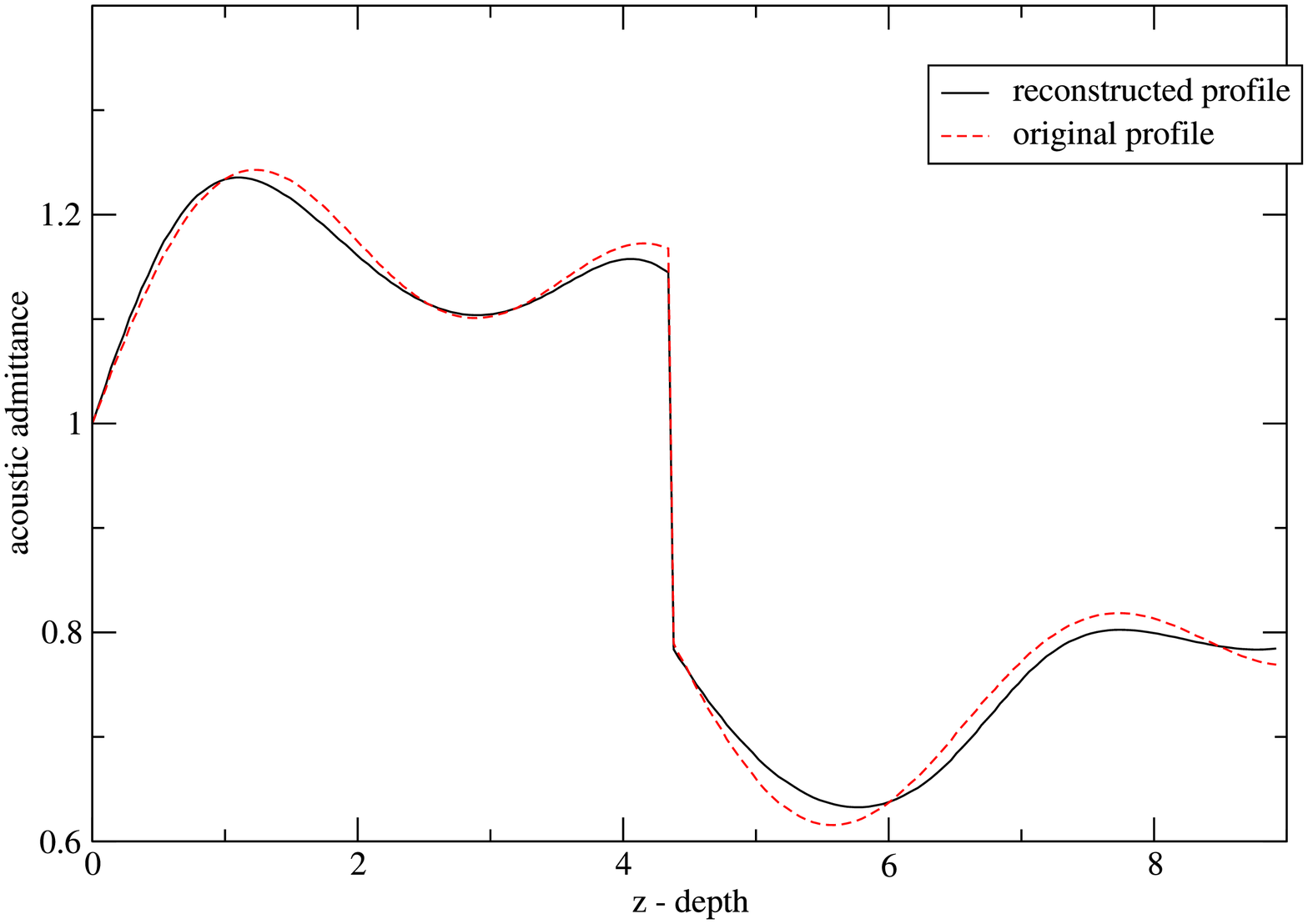}}
\put(210,305){(b)}
\put(207,120){\includegraphics[scale=0.30]{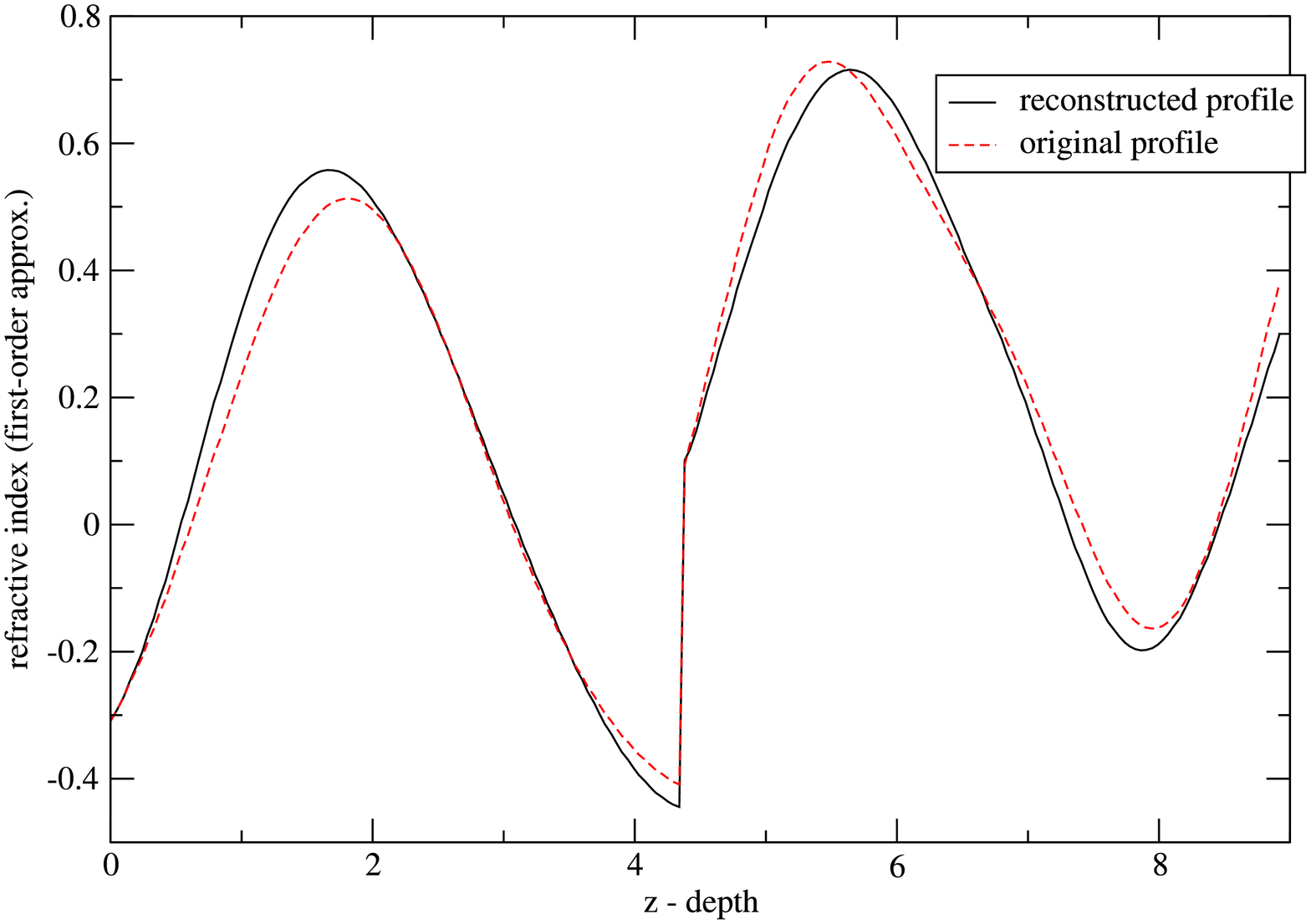}}
\end{picture}
\vspace{-5cm}
\caption{ Numerical values of the acoustic admittance $\sigma_0$ - (a) and $n_1$ - (b) against original profile  with $p_0=1, p_1=0.17, p_2=-0.035, q=0.1, f_0=1.7$, $r_0=0, r_1=-0.28,r_2=-0.03, q_1=0.37, q_2=-0.04, f_1=1.2$ before the discontinuity, and $p_0=0.8, p_1=-0.1, p_2=0.023, q=-0.1, f_0=1.5$, $r_0=0.2, r_1=-0.13, r_2=-0.03, q_1=0.25, q_2=0.04, f_1=1.5$ behind the discontinuity.}
\end{figure}

\begin{figure}[t]
\begin{picture}(400,320)
\put(-20,305){(a)}
\put(-20,120){\includegraphics[scale=0.30]{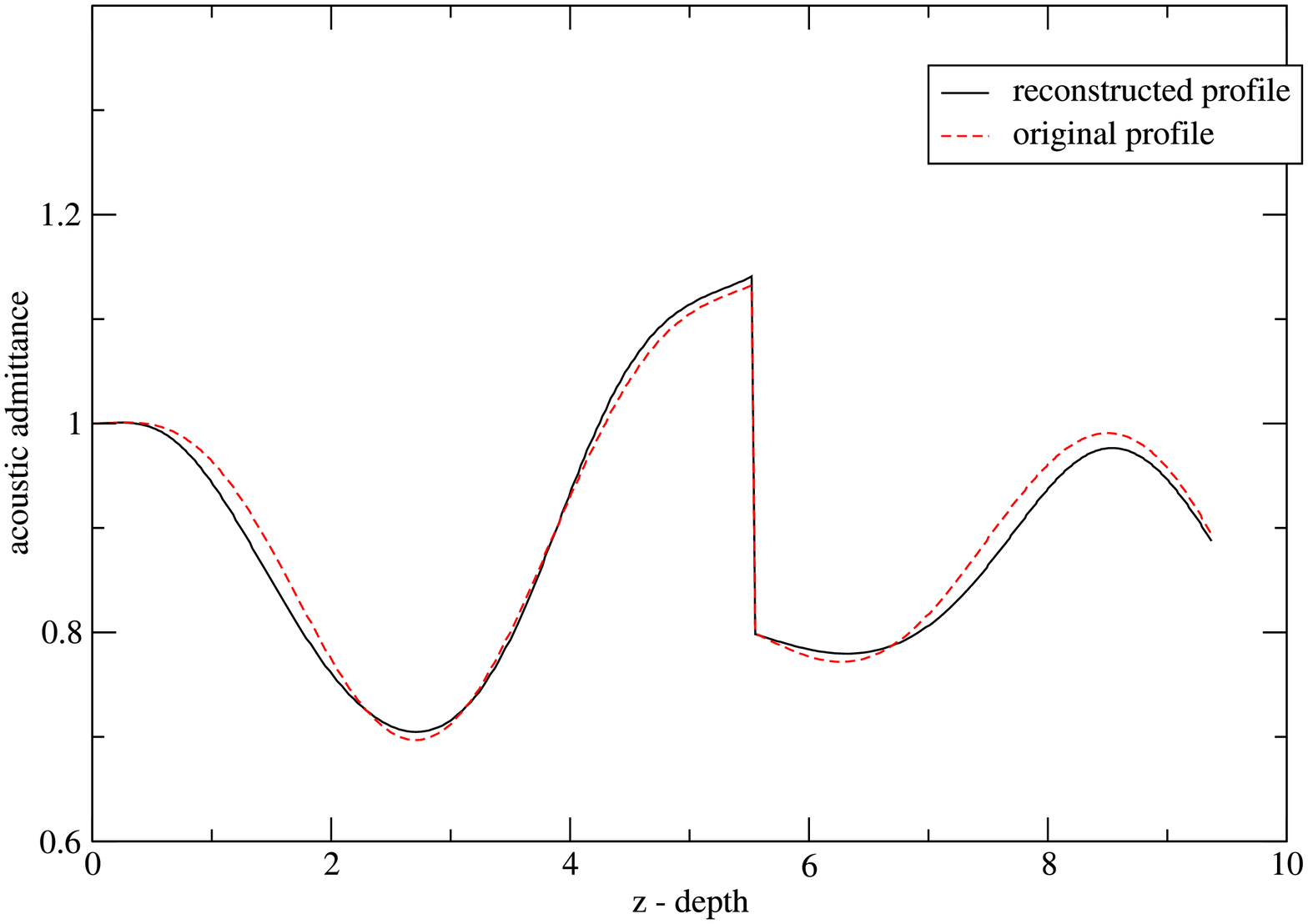}}
\put(210,305){(b)}
\put(207,120){\includegraphics[scale=0.30]{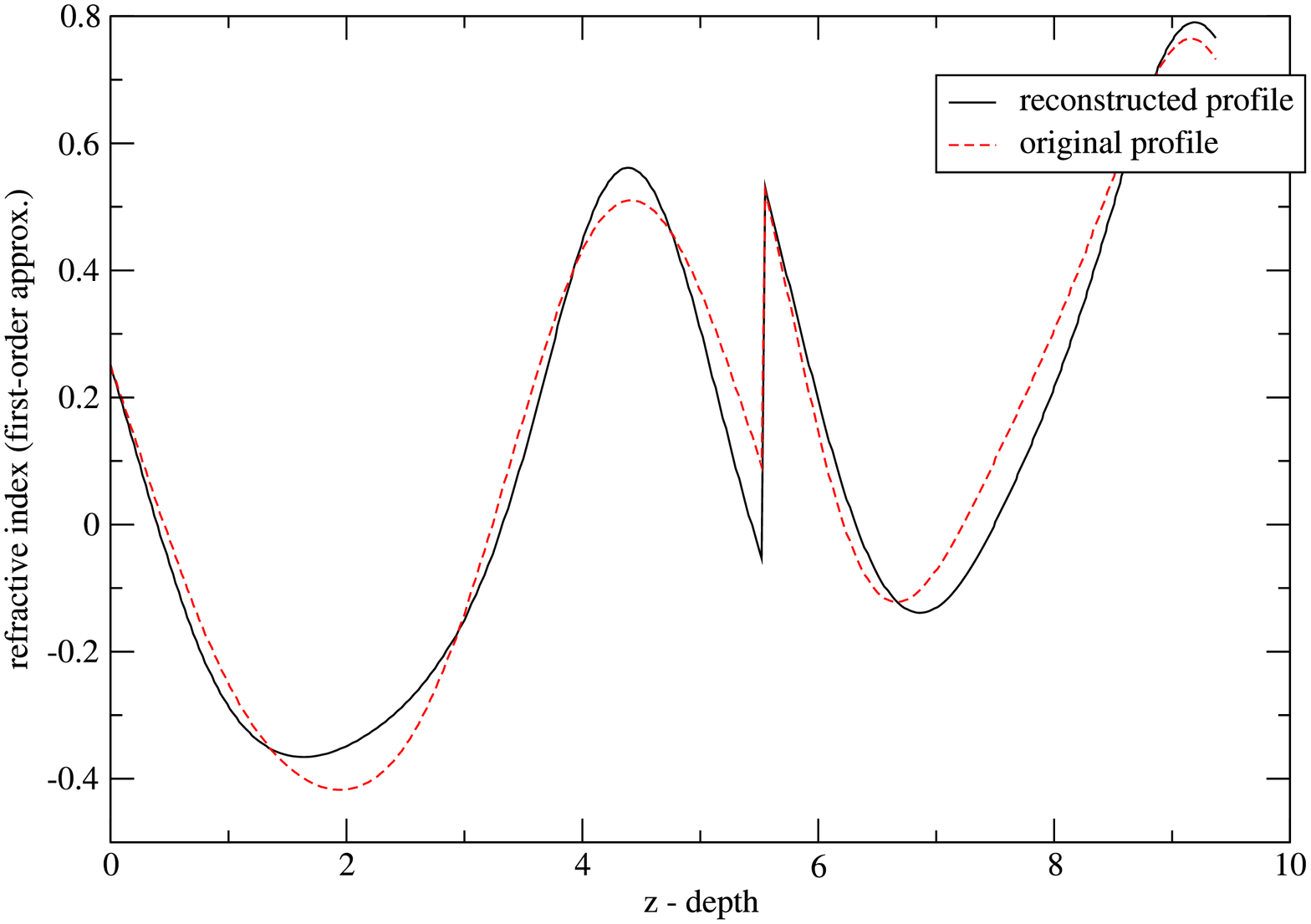}}
\end{picture}
\vspace{-5cm}
\caption{ Numerical values of the acoustic admittance $\sigma_0$ - (a) and $n_1$ - (b) against original profile  with $p_0=1, p_1=-0.17, p_2=0.035, q=0.1, f_0=1.7$, $r_0=0, r_1=0.28, r_2=-0.03, q_1=-0.37, q_2=-0.04, f_1=1.2$ before the discontinuity, and $p_0=0.8, p_1=0.1, p_2=-0.023, q=-0.1, f_0=1.7$ , $r_0=0.2, r_1=0.13, r_2=0.03, q_1=-0.25, q_2=-0.04, f_1=1.5$ behind the discontinuity.}
\end{figure}

\begin{figure}[t]
\begin{picture}(400,320)
\put(-20,305){(a)}
\put(-30,120){\includegraphics[scale=0.30]{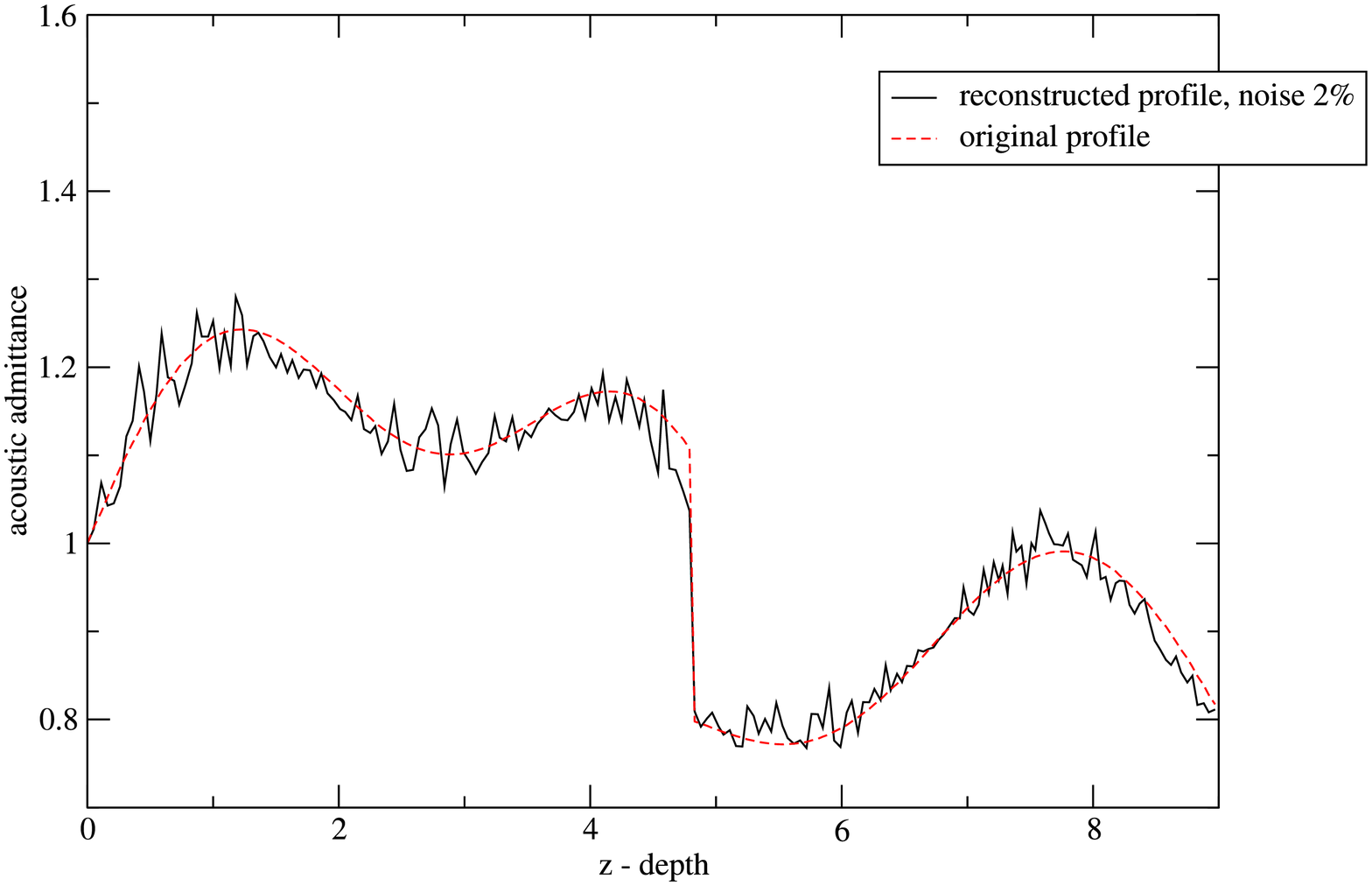}}
\put(210,305){(b)}
\put(207,120){\includegraphics[scale=0.30]{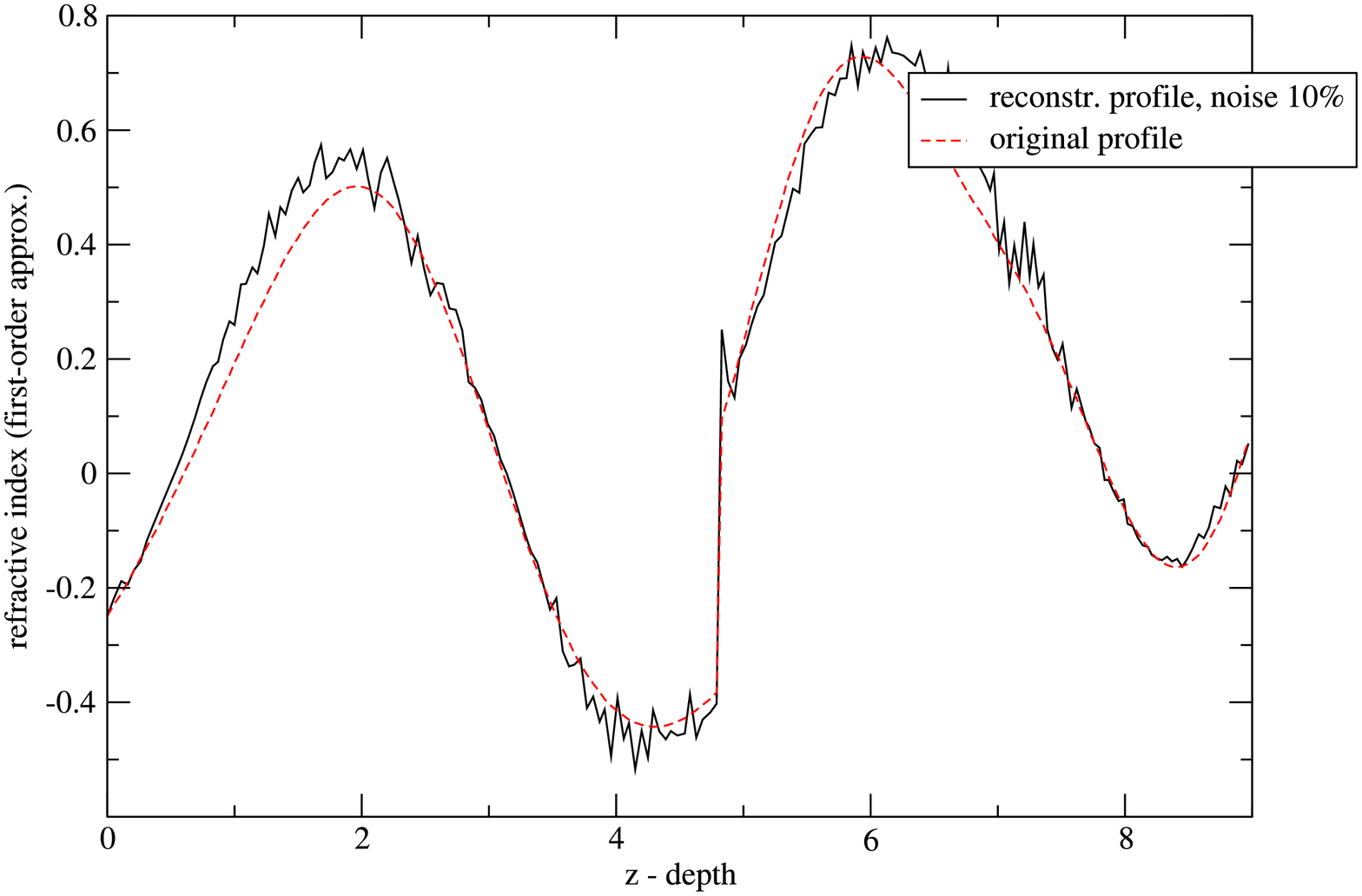}}
\end{picture}
\vspace{-5cm}
\caption{Numerical values of the acoustic admittance $\sigma_0$ - (a) and $n_1$ - (b) against original profile  with $p_0=1, p_1=0.17, p_2=-0.035, q=0.1, f_0=1.7$, $r_0=0, r_1=-0.28, r_2=0.03, q_1=0.37, q_2=-0.04, f_1=1.2$ before the discontinuity, and $p_0=0.8, p_1=0.1, p_2=-0.023, q=-0.1, f_0=1.7$ , $r_0=0.2, r_1=-0.13, r_2=-0.03, q_1=0.25, q_2=0.04, f_1=1.5$ behind the discontinuity  in the case the response data were corrupted by noise - $2 \%$ for $R_0$ and $10 \%$ for $R_1$.}
\end{figure}

\begin{figure}[t]
\begin{picture}(400,320)
\put(-20,305){(a)}
\put(-30,120){\includegraphics[scale=0.30]{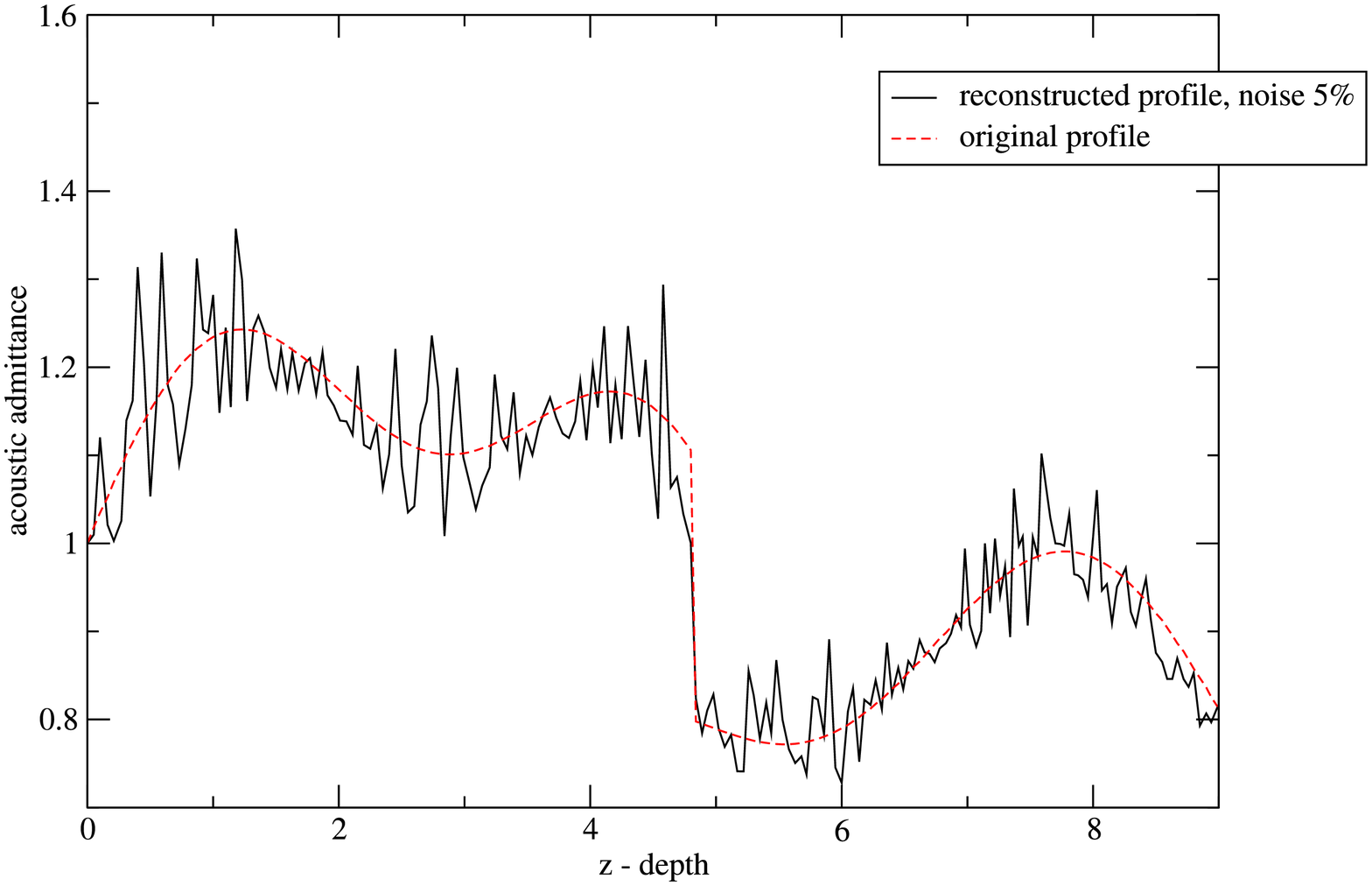}}
\put(210,310){(b)}
\put(207,120){\includegraphics[scale=0.30]{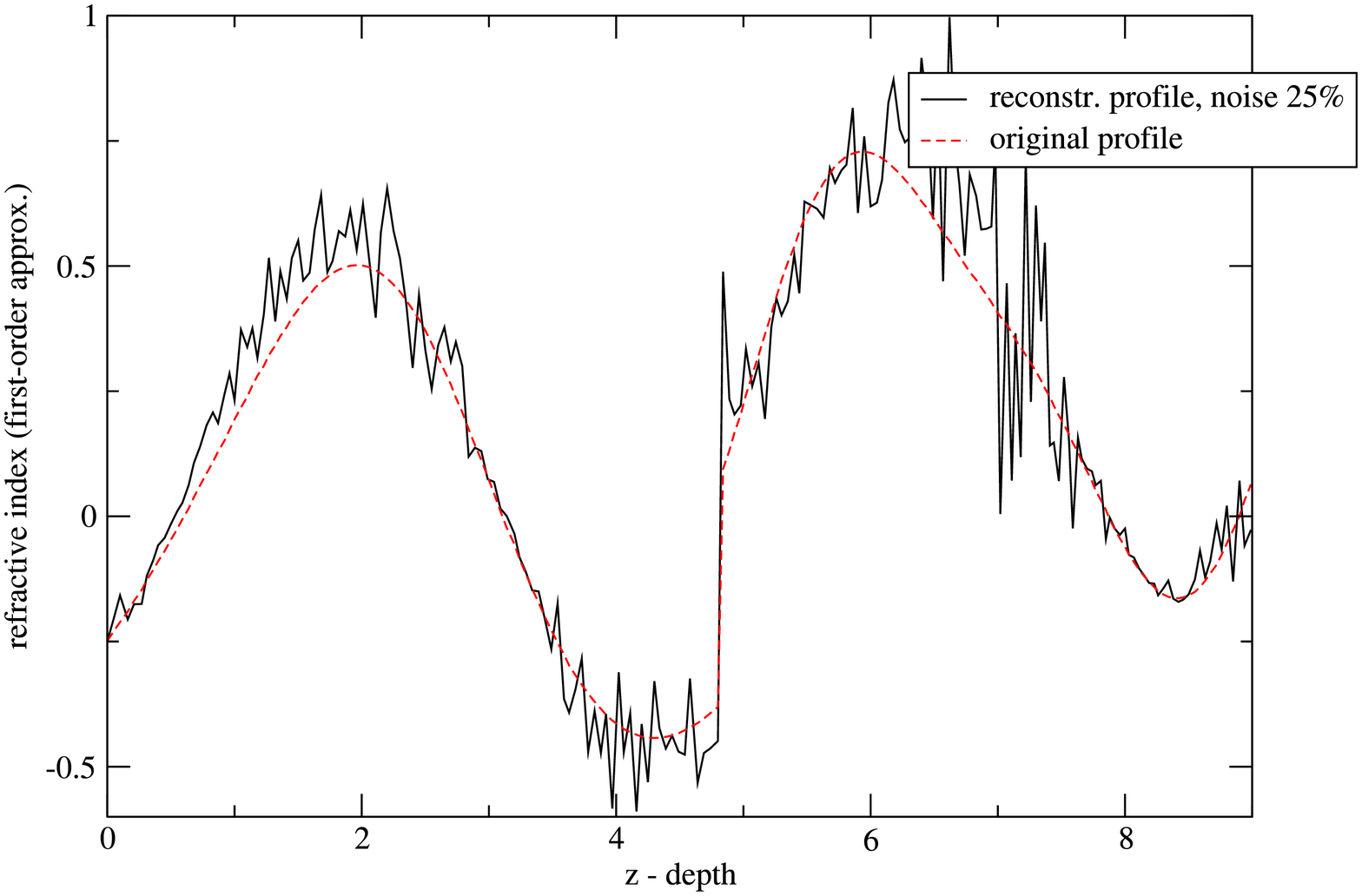}}
\end{picture}
\vspace{-5cm}
\caption{Numerical values of the acoustic admittance $\sigma_0$ - (a) and $n_1$ - (b) against original profile  with $p_0=1, p_1=0.17, p_2=-0.035, q=0.1, f_0=1.7$, $r_0=0, r_1=-0.28, r_2=0.03, q_1=0.37, q_2=-0.04, f_1=1.2$ before the discontinuity, and $p_0=0.8, p_1=0.1, p_2=-0.023, q=-0.1, f_0=1.7$ , $r_0=0.2, r_1=-0.13, r_2=-0.03, q_1=0.25, q_2=0.04, f_1=1.5$ behind the discontinuity  in the case the response data were corrupted by noise - $5 \%$ for $R_0$ and $25 \%$ for $R_1$.}
\end{figure}

\begin{figure}[t]
\begin{picture}(400,320)
\put(-20,305){(a)}
\put(-30,120){\includegraphics[scale=0.30]{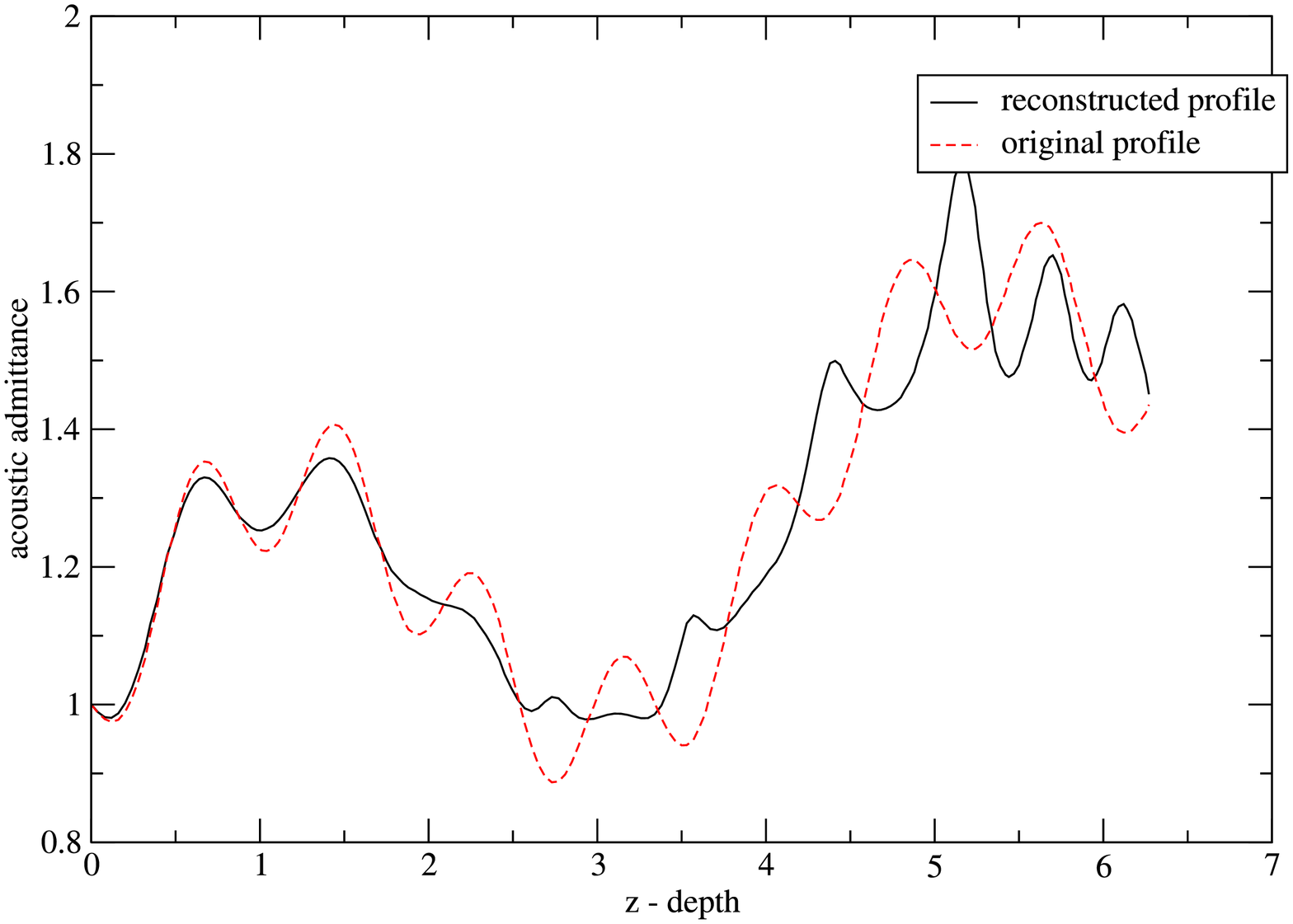}}
\put(210,310){(b)}
\put(207,120){\includegraphics[scale=0.30]{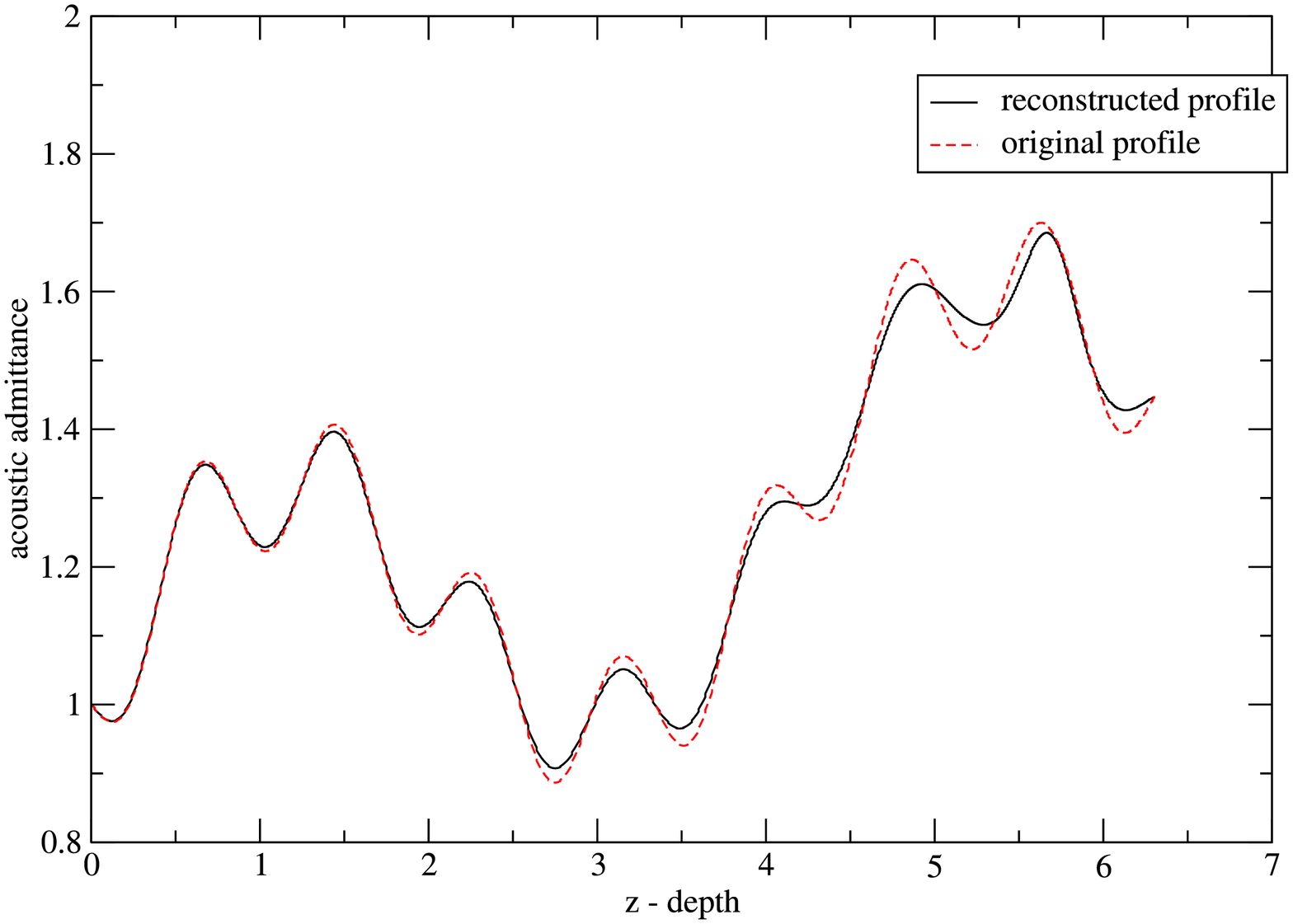}}
\end{picture}
\vspace{-5cm}
\caption{Numerical values of the acoustic admittance $\sigma_0$  against original profile  given by $\sigma_0=1+0.07z+0.25\sin1.5z-0.1\sin7.5z$ with $\delta y=0.04$ - (a)   and $\delta y=0.008$ - (b).}
\end{figure}

\newpage

\section{Concluding remarks} 
\subsection{}
Typical distances of interest in seismology/oil exploration are few kilometers. As a typical velocity 
of the wave propagation is around $2-2.5 km/sec$,  in the travel-time coordinates, $x,z=O(1)$.
 As we have already mentioned in Introduction, the method described in the paper can, in principle,
 reconstruct the velocity profile in this region up to an error of the order $O(\e^2)$. Methods 
 based on an approximation of WLIM by a purely layered medium would give rize to an error
 of the order $O(\e^2+\e|x|)$. A natural way to improve the result when using inversion techniques
 for purely layered medium is to increase the number of sources placing them at  distance
 $O(\e)$. However, in applications to seismology/oil exploration this is not always possible. Indeed,
 a typical structure of the earth contains, in addition to WLIM, various inclusion of different nature 
 with domains of interest often lying below these inclusions. In map migration method, the rays used often propagate oblique to the surface $y=0$ with their substantial part lying in WLIM making 
 it desirable to know well the properties of this medium. Taking into account that, in order to determine
 the velocity profile in WLIM up to depth $T=O(1)$ it is necessary to make measurements during the time
 interval $0<t<2T$, the sources should be located at a distance $O(1)$ from the inclusion 
 not to be contaminated by its influence. Therefore, using inversion methods based 
 on an approximation by a purely layered medium,  we would end up with a reconstruction error,
 near inclusion, of the order $O(\e)$.
 
 \subsection{}
 Another observation, partly related to the above one, concerns with the case when it is 
 necessary/desirable to make measurements only on a part of the ground surface, $y=0$, near  
 the origin. Observe that, although  integrals  (\ref{26.15}),
 (\ref{26.16}) is taken over $R^2$, due to the finite velocity of the wave propagation,
 $R(x,t)=0$ for $x$ with $d(x,0)>t$, where $d((x,y),\,({\tilde x}, {\tilde y})$
 are the distance in the metric $dl^2=c^{-2}dx^2+dy^2$. Consider e.g. an inclusion 
 located near $y=0$ at the distance less then $2T$ from the origin  which would contaminate
  the measurements.  If we, however, make measurements near the origin, the inclusions 
  starts to affect our measurements only when distance goes down to $T$. According to a result
  obtained by the BC-method and valid for a general multidimensional medium, making
  measurements on a subdomain, $\Gamma$, on the surface during time $2T$ makes possible,
  in principle, to recover the velocity profile in the $T-$neighbourhood of $\Gamma$ \cite{KKL2}.
  In the case of a layered medium, due to \cite{Rak} it is even sufficient to make measurements 
  in a single point on $y=0$. 
  
  Let us show that a simple modification of the procedure described in the paper makes it
  possible to determine $c_0(y),\, c_1(y),\, 0<y<T$ given $R(x,t)$ for $|x| <2a,\, t <2T$, where
  $a>0$ is arbitrary. Denote by $\hat c=\max c(x,y),$ over $(x,y)$ lying at the distance less than
  $2T$ from the origin. Observe that, for any $b>0$, the layer $y>b$ affects $R(x,t)$ with
  $|x|>2a$ only when $t >2t(a,b)$. Here $2t(a,b)$ is the time needed for a wave from the origin
  which propagates through a medium with the length element $dl^2= \hat c^{-2} dx^2+dy^2$ to reach the layer $y=b$ and return to the surface $y=0$ at a point $|x|>2a$, i.e.
 $$
 t(a,b)= \sqrt{b^2+a^2 \hat c^{-2}}.
 $$
 Therefore, if we know $c_0(y),\,c_1(y)$ for $y<b$ and $R(x,t)$ for $|x|<2a, \, t<2T$,
 we can determine, up to an error of the order $O(\e^2)$,  $R(x,t)$ for all $x \in R^2$ 
 and $t <2t(a,b)$. Clearly, the procedure described makes it possible to determine
 $c_0(y),\,c_1(y)$ for $y<t(a,b)$. Iterating this process, we reach the level $y=T$ in a finite
 number of steps.

\bigskip
{\bf Acknowledgements}
The authors would like to acknowledge the financial support from EPSRC grants GR/R935821/01 and GR/S79664/01. they are grateful to Prof. C. Chapman for numerous consultations on the
geophysical background of the problem,
Dr. K. Peat for the assistance with numerics for the direct problem and Prof. A.P.Katchalov for
stimulating discussions.


\end{document}